\documentclass[11pt]{article}
\usepackage{amsmath,amssymb,graphicx,epsfig}
\usepackage[margin=0.8in]{geometry}

\newcommand{\D}{{\mathrm{d}}}

\begin{document}

\title{Dynamical modeling of microRNA action on the protein translation process}
\author{Andrei Zinovyev $^{1,2,3,6,*}$, Nadya Morozova$^{4}$, Nora Nonne$^{4}$ \\ Emmanuel Barillot$^{1,2,3}$, Annick Harel-Bellan$^{4}$, Alexander N. Gorban$^{5,6}$
\\\\ {\small$^1$ Institut Curie, Service Bioinformatique,
Paris,France}\\ {\small$^2$ INSERM, U900, Paris, F-75248 France}\\
{\small $^3$ Ecole des Mines de Paris, ParisTech, Fontainebleau,
France}\\ {\small $^4$ CNRS FRE 2944, Institut Andr\'{e} Lwoff,
Villejuif, France}\\ {\small $^5$ University of Leicester, UK}\\
{\small $^6$ Institute of Computational Modeling SB RAS,
Krasnoyarsk, Russia}\\ {\small * Correspondent author e-mail:
andrei.zinovyev@curie.fr }}

\maketitle

\abstract{ Protein translation is a multistep process which can be
represented as a cascade of biochemical reactions (initiation,
ribosome assembly, elongation, etc.), the rate of which can be
regulated by small non-coding microRNAs through multiple
mechanisms. It remains unclear what mechanisms of microRNA action
are most dominant: moreover, many experimental reports deliver
controversal messages on what is the concrete mechanism actually
observed in the experiment. Parker and Nissan \cite{Nissan2008}
demonstrated that it is impossible to distinguish alternative
biological hypotheses using the steady state data on the rate of
protein synthesis. For their analysis they used two simple kinetic
models of protein translation. In contrary, we show that dynamical
data allow to discriminate some of the mechanisms of microRNA
action. We demonstrate this using the same models as in
\cite{Nissan2008} for the sake of comparison but the methods
developed (asymptotology of biochemical networks) can be used for
other models. As one of the results of our analysis, we formulate
a hypothesis that the effect of microRNA action is measurable and
observable only if it affects the dominant system (generalization
of the limiting step notion for complex networks) of the protein
translation machinery. The dominant system can vary in different
experimental conditions that can partially explain the existing
controversy of some of the experimental data.}
\date{}

\section{Introduction}

MicroRNAs (miRNAs) are currently considered as key regulators of a
wide variety of  biological pathways, including development,
differentiation and oncogenesis. Recently, remarkable progress was
made in the understanding of microRNA biogenesis, function and
mechanism of action. Mature microRNAs are incorporated into the
RISC effector complex, which includes as a key component an
Argonaute protein. MicroRNAs affect gene expression by guiding the
RISC complex toward specific target mRNAs. The exact mechanism of
this inhibition is still a matter of debate. In the past few
years, several mechanisms have been reported, some of which are
contradictory (for review, see \cite{Chekulaeva2009,Eulalio2008,
Filipowicz2008}). These include in particular inhibition of
translation initiation (acting at the level of cap-40S or
40S-AUG-60S association steps), inhibition of translation
elongation or premature termination of translation.
MicroRNA-mediated mRNA decay and sequestration of target mRNAs in
P-bodies have been also proposed. Moreover, some microRNAs mediate
target mRNA cleavage \cite{Yekta2004}, chromatin reorganization
followed by transcriptional repression or activation
\cite{Kim2008, Place2008}, or translational activation
\cite{Orom2008, Vasudevan2007}.

\begin{figure}
\centerline{
\includegraphics[width=10cm]{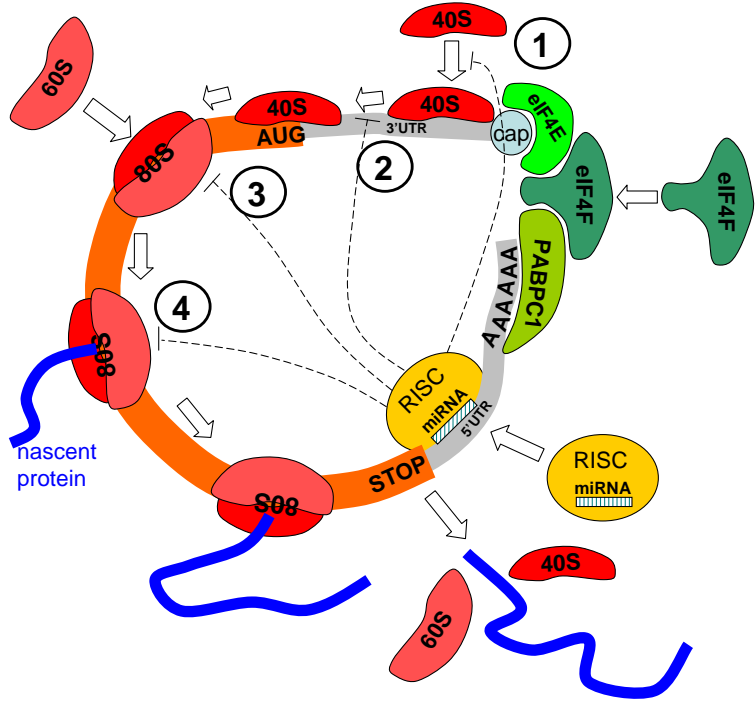}
} \caption{\label{translation_miRNA} Interaction of microRNA with
protein translation process. Four mechanisms of translation
repression which are considered in the mathematical modeling are
indicated: 1) on the initiation process, preventing assembling of
the initiation complex; 2) on a late initiation step, such as
searching for the start codon; 3) on the ribosome assembly; 4) on
the translation process. There exist other mechanisms of microRNA
action on protein translation (transcriptional, transport to
P-bodies, ribosome drop-off, co-translational protein degradation
and others) that are not considered in this paper. Here 40S and
60S are light and heavy components of the ribosome, 80S is the
assembled ribosome bound to mRNA, eIF4F is an translation
initiation factor, PABC1 is the Poly-A binding protein, ``cap'' is
the mRNA cap structure needed for mRNA circularization, RISC is
the RNA-induced silencing complex. }
\end{figure}

The most frequently reported, but also much debated, mechanism of
gene repression by microRNAs occurs at the level of mRNA
translation. This is particularly true when the microRNA is not
fully complementary to its target. At this level, several mode of
actions have been suggested (see Fig.~\ref{translation_miRNA}).
Historically, the first proposed mechanism was inhibition of
translation elongation. The major argument supporting this
hypothesis was the observation that the inhibited mRNA remained
associated with the polysomal fraction (in which mRNAs are
associated with polysomes), an observation that was reproduced in
different systems \cite{Gu2009, Nottrott2006, Petersen2006,
Olsen99, Maroney2006}. The idea of a post-initiation mechanism was
further supported by the observation that some mRNAs can be
repressed by a microRNA even when their translation is
cap-independent (mRNAs with an IRES or A-capped)
\cite{Baillat2009, Karaa2009, Lytle2007, Petersen2006}. Although
it was initially proposed that the ribosomes was somehow ``frozen"
on the mRNA, it is important to note that it is difficult to
discriminate experimentally between different post-initiation
potential mechanisms, such as elongation inhibition, premature
ribosome dissociation (``ribosome drop-off") or normal elongation
but nascent polypeptide degradation. The last proposition (which
can occur in conjonction with the two others) is supported by the
fact that the mRNA-polysomal association is puromycin-sensitive,
indicating that it depends on a peptidyl-transferase activity
\cite{Bandres2006, Maroney2006}. However, no nascent peptide has
ever been experimentally demonstrated; thus degradation would
occur extremely rapidly after synthesis \cite{Pillai2005,
Petersen2006, Nottrott2006}. Degradation, in any case, is
proteasome-independent \cite{Pillai2005}. Premature ribosome
dissociation is supported by decreased read-through of inhibited
mRNA \cite{Petersen2006}. Ribosome drop-off and/or ribosomal
``slowing" are supported by the slight decrease in the number of
associated ribosomes observed in some studies \cite{Nottrott2006,
Maroney2006}.

Concurrently, several reports have been published indicating an
action of microRNAs at the level of initiation. An increasing
number of papers reports that microARN-targetted mRNAs shift
toward light fractions in polysomal profiles
\cite{Bhattacharyya2006, Kiriakidou2007, Pillai2005}. This shows a
decrease of mature translating ribosomes, suggesting that
microRNAs act on the initiation step. Moreover, several reports
show that microRNA-mediated inhibition is relieved when
translation is driven by a cap-independent mechanism such as
IRES-mRNA or A-capped-mRNA \cite{Humphreys2005, Kiriakidou2007,
Pillai2005}. This observation was confirmed in several in-vitro
studies \cite{Mathonnet2007, Thermann2007, Wang2006,
Wakiyama2007}. In particular, in one of those, an excess of eIF4F
could relieve the inhibition, and inhibition lead to decreased 80S
in polysomal gradient \cite{Mathonnet2007}.

Most of the data indicating a shift toward light polysomal
fraction or requirement for a cap-dependent translation are often
interpreted in favour of involvement of microRNAs at early steps
of translation, i.e., cap binding and 40S recruitment. However,
some of them are also compatible with a block at the level of 60S
subunit joining. This hypothesis is also supported by in-vitro
experiments showing a lower amount of 60S relative to 40S on
inhibited mRNAs. Moreover, toe-printing experiments show that 40S
is positioned on the AUG \cite{Wang2008}. Independently, it was
shown that eIF6, an inhibitor of 60S joining, is required for
microRNA action \cite{Chendrimada2007}, but this was contradicted
by other studies \cite{Eulalio2008}.

Thus, data regarding the exact step of translational inhibition
are clearly contradictory.  Taking also into account the data
about mRNA degradation and P-bodies localization, it is difficult
to draw a clear picture of the situation, and the exact mechanism
by which microRNA repress mRNA expression is highly controversial,
let alone the interrelations between the different mechanisms and
their possible concomitant action. Several attempts to integrate
the different hypothesis have been made \cite{Chekulaeva2009,
Eulalio2008, Filipowicz2008, Jackson2007,Kozak2008,
Valencia-Sanchez2006}. For example, one mechanism could act as a
``primary" effect, and the other as a ``secondary" mechanism,
either used to reinforce the inhibition or as back-up mechanism.
In others, the different mechanisms could all coexist, but occur
differentially depending on some yet unidentified characteristic.
For example, it has been observed than the same mRNA targeted by
the same microRNA can be regulated either at the initiation or the
elongation step depending on the mRNA promoter and thus on mRNA
nuclear history \cite{Kong2008}. It was also proposed that
technical (experimental) problems, including the variety of
experimental systems used, may also account for these
discrepancies \cite{Chekulaeva2009, Eulalio2008, Filipowicz2008}.
However, this possibility does not seem to be sufficient to
provide a simple and convincing explanation to the reported
discrepancies.

A possible solution to exploit the experimental observations and
to provide a rational and straightforward interpretation is the
use of mathematical models for microRNA action. Nissan and Parker
recently studied the steady states of two simple kinetic models
\cite{Nissan2008} and provided critical analysis for the
experiments with alternative mRNA cap structures and IRES elements
\cite{Mathonnet2007,Thermann2007,Wakiyama2007}, leading to
possible explanation of the conflicting results. The authors
suggested that the relief of translational repression upon
replacement of the cap structure can be explained if microRNA is
acting on a step which is not rate-limiting in the modified
system, in which case, an effect of microRNA can simply not be
observed. It was claimed that it is impossible to discriminate
between two alternative interpretations of the biological
experiments with cap structure replacement from the sole
monitoring of the steady state level of protein \cite{Nissan2008}.

Two remarks can be made in this regard. First, in practice not
only the steady state level of protein can be observed but also
other dynamical characteristics, such as the {\it relaxation
time}, i.e. the time needed to achieve the steady state rate after
a perturbation (such as restarting the translation process). We
argue that having these measurements in hands, {\it one can
distinguish between two alternative interpretations}. In this
paper we provide such a method from the same models as constructed
by Nissan and Parker, for comparison purposes. However, the method
applied can be easily generalized for other models.

Second, even in the simple non-linear model of protein
translation, taking into account the recycling of ribosomal
components, it is difficult to define what is the rate limiting
step. It is known from the theory of asymptotology of biochemical
networks \cite{Gorban2009} that even in complex linear systems the
``rate limiting place'' notion is not trivial and can not be
reduced to a single reaction step. Moreover, in non-linear systems
the ``rate limiting place'' can change with time and depend on the
initial conditions. Hence, conclusions of \cite{Nissan2008} should
be re-considered for the non-linear model, made more precise and
general. The notion of rate limiting step should be replaced by
the notion of {\it dominant system}.

In this paper we perform careful analysis of the Nissan and
Parker's models and provide their approximate analytical
solutions, which allows us to generalize the conclusions of
\cite{Nissan2008} and make new testable predictions on the
identifiability of active mechanism of microRNA-dependent protein
translation inhibition.

\section{Methods}

\subsection{Asymptotology and dynamical limitation theory for biochemical reaction networks}

Most of mathematical models that really work are simplifications
of the basic theoretical models and use in the backgrounds an
assumption that some terms are big, and some other terms are small
enough to neglect or almost neglect them. The closer consideration
shows that such a simple separation on ``small" and ``big" terms
should be used with precautions, and special culture was
developed. The name ``asymptotology" for this direction of science
was proposed by \cite{Kruskal62} defined as ``the art of handling
applied mathematical systems in limiting cases".

In chemical kinetics three fundamental ideas were developed for
model simplification: {\em quasiequilibrium asymptotic} (QE), {\em
quasi steady-state asymptotic} (QSS) and the idea of {\em limiting
step}.

In the IUPAC Compendium of Chemical Terminology (2007) one can
find a definition of limiting step \cite{R-cont}: ``A
rate-controlling (rate-determining or rate-limiting) step in a
reaction occurring by a composite reaction sequence is an
elementary reaction the rate constant for which exerts a strong
effect -- stronger than that of any other rate constant -- on the
overall rate."

Usually when people are talking about limiting step they expect
significantly more: there exists  a rate constant which exerts
such a strong effect on the overall rate that the effect of all
other rate constants together is significantly smaller. For the
IUPAC Compendium definition a rate-controlling step always exists,
because among the control functions generically exists the biggest
one. On the contrary,  for the notion of limiting step that is
used in practice, there exists a difference between systems with
limiting step and systems without limiting step.

During XX century, the concept of the limiting step was revised
several times. First simple idea of a ``narrow place" (the least
conductive step) could be applied without adaptation only to a
simple cycle or a chain of irreversible steps that are of the
first order (see Chap. 16 of the book \cite{Johnston} or the paper
by \cite{Boyd} or the section \ref{cycleSection} of this paper).
When researchers try to apply this idea in more general situations
they meet various difficulties.

Recently, we proposed a new theory of dynamic and static
limitation in multiscale reaction networks \cite{GorbanRadul2008,
Gorban2009}. This approach allows to find the simplest network
which can substitute a multiscale reaction network such that the
dynamics of the complex network can be approximated by the simpler
one. Following the asymptotology terminology \cite{White2006}, we
call this simple network the {\it dominant system} (DS). In the
simplest cases, the dominant system is a subsystem of the original
model. However, in the general case, it also includes new
reactions with kinetic rates expressed through the parameters of
the original model, and rates of some reactions are renormalized:
hence, {\it in general, the dominant system is not a subsystem of
the original model}.

The dominant systems can be used for direct computation of steady
states and relaxation dynamics, especially when kinetic
information is incomplete, for design of experiments and mining of
experimental data, and could serve as a robust first approximation
in perturbation theory or for preconditioning. Dominant systems
serve as correct generalization of the limiting step notion in the
case of complex multiscale networks. They can be used to answer an
important question: given a network model, which are its critical
parameters? Many of the parameters of the initial model are no
longer present in the dominant system: these parameters are
non-critical. Parameters of dominant systems indicate putative
targets to change the behavior of the large network.

Most of reaction networks are nonlinear, it is nevertheless useful
to have an efficient algorithm for solving linear problems. First,
nonlinear systems often include linear subsystems, containing
reactions that are (pseudo)monomolecular with respect to species
internal to the subsystem (at most one internal species is
reactant and at most one is product). Second, for binary reactions
$A + B \to ...$, if concentrations of species $A$ and $B$ ($c_A,
c_B$) are well separated, say $c_A \gg c_B$ then we can consider
this reaction as $B \to ...$ with rate constant proportional to
$c_A$ which is practically constant, because its relative changes
are small in comparison to relative changes of $c_B$. We can
assume that this condition is satisfied for all but a small
fraction of genuinely non-linear reactions (the set of non-linear
reactions changes in time but remains small). Under such an
assumption,  non-linear behavior can be approximated as a sequence
of such systems, followed one each other in a sequence of ``phase
transitions". In these transitions, the order relation between
some of species concentrations changes. Some applications of this
approach to systems biology are presented by
\cite{RadGorZinLil2008}. The idea of controllable linearization
``by excess" of some reagents is in the background of the
efficient experimental technique of Temporal Analysis of Products
(TAP), which allows to decipher detailed mechanisms of catalytic
reactions \cite{TAP}.

Below we give some details on the approaches used in this paper to
analyse the models by Nissan and Parker \cite{Nissan2008}.

\subsection{Notations}

To define a chemical reaction network, we have to introduce:
\begin{itemize}
\item{a list of components (species);}
\item{a list of elementary reactions;}
\item{a kinetic law of elementary reactions.}
\end{itemize}
The list of components is just a list of symbols (labels) $A_1,...
A_n$. Each elementary reaction is represented by its {\it
stoichiometric equation}

\begin{equation}
\sum_{si} \alpha_{si} A_i \to \sum_{si} \beta_i A_i ,
\end{equation}

\noindent where $s$ enumerates the elementary reactions, and the
non-negative integers $\alpha_{si}$, $\beta_{si}$ are the {\it
stoichiometric coefficients}.  A stoichiomentric vector $\gamma_s$
with coordinates $\gamma_{si}=\beta_{si}-\alpha_{si}$ is
associated with each elementary reaction.

A non-negative real {\it extensive} variable $N_i \geq 0$, amount
of $A_i$, is associated with each component $A_i$. It measures
``the number of particles of that species" (in particles, or in
moles). The concentration of $A_i$ is an {\it intensive} variable:
$c_i = N_i/V$, where $V$ is volume. In this paper we consider the
volume (of cytoplasm) to be constant. Then the kinetic equations
have the following form

\begin{equation}\label{ConcKinur}
\frac{\D c}{\D t}=\sum_s w_s(c,T)\gamma_s + \upsilon,
\end{equation}

\noindent where $T$ is the temperature, $w_s$ is the rate of the
reaction $s$, $\upsilon$ is the vector of external fluxes
normalized to unite volume. It may be useful to represent external
fluxes as elementary reactions by introduction of new component
$\varnothing$ together with income and outgoing reactions
$\varnothing \to A_i$ and $A_i \to \varnothing$.

The most popular {\it kinetic law} of elementary reactions is the
{\it mass action law} for perfect systems:
\begin{equation}\label{MAL}
w_s(c,T)=k_s\prod c_i^{\alpha_{si}},
\end{equation}
where $k_s$ is a ``kinetic constant" of the reaction $s$.

\subsection{Quasy steady-state and quasiequilibrium
asymptotics}\label{asymptoticApprox}

Quasiequilibrium approximation uses the assumption that a group of
reactions is much faster then other and goes fast to its
equilibrium. We use below superscripts `$^{\rm f}$' and `$^{\rm
s}$' to distinguish fast and slow reactions. A small parameter
appears in the following form
\begin{equation}\label{ConcKinurQE}
\begin{split}
\frac{\D c}{\D t}= &\sum_{s, \ {\rm slow}} w_s^{\rm
s}(c,T)\gamma_s^{\rm s} + \frac{1}{\varepsilon} \sum_{\varsigma, \
{\rm fast}} w^{\rm f}_{\varsigma}(c,T)\gamma_{\varsigma}^{\rm f},
\end{split}
\end{equation}
To separate variables, we have to study the spaces of linear
conservation law of the initial system (\ref{ConcKinurQE}) and of
the fast subsystem
 $$\frac{\D c}{\D t}=\frac{1}{\varepsilon} \sum_{\varsigma, \ {\rm
 fast}} w^{\rm f}_{\varsigma}(c,T)\gamma_{\varsigma}^{\rm f}$$
If they coincide, then the fast subsystem just dominates, and
there is no fast-slow separation for variables (all variables are
either fast or constant). But if there exist additional linearly
independent linear conservation laws for the fast system, then let
us introduce new variables: linear functions $b^1(c),... b^n(c)$,
where $b^1(c),... b^m(c)$ is the basis of the linear conservation
laws for the initial system, and $b^{1}(c),... b^{m+l}(c)$ is the
basis of the linear conservation laws for the fast subsystem. Then
$b^{m+l+1}(c),... b^n (c)$ are fast variables, $b^{m+1}(c),...
b^{m+l}(c)$ are slow variables, and $b^{1}(c),... b^{m}(c)$ are
constant. The {\it quasiequilibrium manifold} is given by the
equations $\sum_{\varsigma} w^{\rm
f}_{\varsigma}(c,T)\gamma_{\varsigma}^{\rm f}=0$ and for small
$\varepsilon$ it serves as an approximation to a slow manifold.

The quasi steady-state (or pseudo steady state) assumption was
invented in chemistry for description of systems with radicals or
catalysts. In the most usual version the species are split in two
groups with concentration vectors $c^{\rm s}$ (``slow" or basic
components) and $c^{\rm f}$ (``fast intermediates"). For catalytic
reactions there is additional balance for $c^{\rm f}$, amount of
catalyst, usually it is just a sum $b_{\rm f}=\sum_i c^{\rm f}_i$.
The amount of the fast intermediates is assumed much smaller than
the amount of the basic components, but the reaction rates are of
the same order, or even the same (both intermediates and slow
components participate in the same reactions). This is the source
of a small parameter in the system. Let us scale the
concentrations $c^{\rm f}$ and $c^{\rm s}$ to the compatible
amounts. After that, the fast and slow time appear and we could
write $\dot{c}^{\rm s} = W^{\rm s}(c^{\rm s},c^{\rm f})$,
$\dot{c}^{\rm f} =\frac{1}{\varepsilon} W^{\rm f}(c^{\rm s},c^{\rm
f})$, where $\varepsilon$ is small parameter,  and functions
$W^{\rm s},W^{\rm f}$ are bounded and have bounded derivatives
(are ``of the same order"). We can apply the standard singular
perturbation techniques. If dynamics of fast components under
given values of slow concentrations is stable, then the slow
attractive manifold exists, and its zero approximation is given by
the system of equations $W^{\rm f}(c^{\rm s},c^{\rm f})=0$.

The QE approximation is also extremely popular and useful. It has
simpler dynamical properties (respects thermodynamics, for
example, and gives no critical effects in fast subsystems of
closed systems). Nevertheless, neither radicals in combustion, nor
intermediates in catalytic kinetics are, in general, close to
quasiequilibrium. They are just present in much smaller amount,
and when this amount grows, then the QSS approximation fails.

The simplest demonstration of these two approximation gives the
simple reaction: $S+E\leftrightarrow SE \to P+E$ with reaction
rate constants $k^{\pm}_1$ and $k_2$. The only possible
quasiequilibrium appears when the first equilibrium is fast:
$k^{\pm}_1 = \kappa^{\pm}/\varepsilon$. The corresponding slow
variable is $C^s=c_S+c_{SE}$, $b_E=c_E+c_{SE}=const$. For the QE
manifold we get a quadratic equation
$\frac{k_1^-}{k_1^+}c_{SE}=c_Sc_E= (C^s-c_{SE})(b_E-c_{SE})$. This
equation gives the explicit dependence $c_{SE}(C^s)$, and the slow
equation reads $\dot{C}^s=-k_2 c_{SE}(C^s)$, $C^s+c_P=b_S=const$.

For the QSS approximation of this reaction kinetics, under
assumption $b_E \ll b_S$, we have fast intermediates $E$ and $SE$.
For the QSS manifold there is a linear equation
$k^+_1c_Sc_E-k_1^-c_{SE} - k_2c_{SE}=0$, which gives us the
explicit expression for $c_{SE}(c_S)$: $c_{SE}=k_1^+ c_S
b_E/(k_1^+ c_S+k_1^-+k_2)$ (the standard Michaelis--Menten
formula). The slow kinetics reads $\dot{c}_S=-k_1^+c_S
(b_E-c_{SE}(c_S)) + k_1^-c_{SE}(c_S)$. The difference between the
QSS and the QE in this example is obvious.

The terminology is not rigorous, and often QSS is used for all
singular perturbed systems, and QE is applied only for the
thermodynamic exclusion of fast variables by the maximum entropy
(or minimum of free energy, or extremum of another relevant
thermodynamic function) principle (MaxEnt). This terminological
convention may be convenient. Nevertheless, without any relation
to terminology, the difference between these two types of
introduction of a small parameter is huge. There exists plenty of
generalizations of these approaches, which aim to construct a slow
and (almost) invariant manifold, and to approximate fast motion as
well. The following references can give a first impression about
these methods: Method of Invariant Manifolds (MIM) (\cite{GorKar,
Roussel91}, Method of Invariant Grids (MIG), a discrete analogue
of invariant manifolds (\cite{Grids}), Computational Singular
Perturbations (CSP) (\cite{Lam1993,LamGous1994,ZaKapers})
Intrinsic Low-Dimensional Manifolds (ILDM)  by \cite{Maas},
developed further in series of works by \cite{BGGMaas2006}),
methods based on the Lyapunov auxiliary theorem
(\cite{KazKraLya}).

\subsection{Multiscale monomolecular reaction networks}

A {\it monomolecular reaction} is such that it has at most one
reactant and at most one product. Let us consider a general
network of monomolecular reactions. This network is represented as
a directed graph (digraph) \cite{Temkin1996}: vertices correspond
to components $A_i$, edges correspond to reactions $A_i \to A_j$
with kinetic constants $k_{ji} > 0$. For each vertex, $A_i$, a
positive real variable $c_i$ (concentration) is defined.
``Pseudo-species" (labeled $\varnothing$) can be defined to
collect all degraded products, and degradation reactions can be
written as $A_i \rightarrow \varnothing$ with constants $k_{0i}$.
Production reactions can be represented as $\varnothing
\rightarrow A_i$ with rates $k_{i0}$. The kinetic equation for the
system is

\begin{equation}\label{kinur}
\frac{\D c_i}{\D t}=k_{i0} + \sum_{j\geq 1}k_{ij} c_j -
\sum_{j\geq 0} k_{ji}c_i,
\end{equation}

\noindent or in vector form: $\dot{c} = K_0+Kc$. Solution of this
system can be reduced to a linear algebra problem: let us find all
left ($l^{i}$) and right ($r^{i}$) eigenvectors of $K$, i.e.:

\begin{equation}\label{eigenvectors}
K r^{i} = \lambda_{i}r^{i}, l^{i} K  = \lambda_{i}l^{i},
\end{equation}

\noindent with the normalisation $(l^{i},r^{i})=\delta_{ij}$,
where $\delta_{ij}$ is Kronecker's delta.  Then the solution of
(\ref{kinur}) is

\begin{equation}\label{kinurSolution}
c(t) = c^s + \sum_{k=1}^{n}r^{k}(l^k,c(0)-c^s)\exp(-\lambda_kt),
\end{equation}

\noindent where $c^s$ is the steady state of the system
(\ref{kinur}), i.e. when all $\frac{\D c_i}{\D t}=0$, and $c(0)$
is the initial condition.

If all reaction constants $k_{ij}$ would be known with precision
then the eigenvalues and the eigenvectors of the kinetic matrix
can be easily calculated by standard numerical techniques.
Furthermore, Singular Value Decomposition (SVD) can be used for
model reduction. But in systems biology models often one has only
approximate or relative values of the constants (information on
which constant is bigger or smaller than another one). Let us
consider the simplest case: when all kinetic constants are very
different (separated), i.e. for any two different pairs of indices
$I = (i, j)$, $J = (i', j')$ we have either $k_I \gg k_J$ or $k_J
\ll k_I$. In this case we say that the system is hierarchical with
timescales (inverses of constants $k_{ij}$, $j \neq 0$) totally
separated.

Linear network with totally separated constants can be represented
as a digraph and a set of orders (integer numbers) associated to
each arc (reaction). The lower the order, the more rapid is the
reaction. It happens that in this case the special structure of
the matrix $K$ (originated from a reaction graph) allows us to
exploit the strong relation between the dynamics (\ref{kinur}) and
the topological properties of the digraph. In this case, possible
values of $l_i$ k are $0$, $1$ and the possible values of $r_i$
are -1, 0, 1 with high precision. In previous works, we provided
an algorithm for finding non-zero components of $l_i$, $r_i$,
based on the network topology and the constants ordering, which
gives a good approximation to the problem solution
\cite{GorbanRadul2008, Gorban2009, RadGorZinLil2008}.

\subsection{Dominant system for a simple
irreversible catalytic cycle with limiting
step}\label{cycleSection}

A linear chain of reactions, $A_1 \to A_2 \to ... A_n$, with
reaction rate constants $k_i$ (for $A_i \to A_{i+1}$), gives the
first example of limiting steps. Let the reaction rate constant
$k_q$ be the smallest one. Then we expect the following behaviour
of the reaction chain in time scale $\gtrsim 1/k_q$: all the
components $A_1,... A_{q-1}$ transform fast into $A_q$, and all
the components $A_{q+1}, ... A_{n-1}$ transform fast into $A_n$,
only two components, $A_q$ and $A_n$ are present (concentrations
of other components are small) , and the whole dynamics in this
time scale can be represented by a single reaction $A_q \to A_n$
with reaction rate constant $k_q$. This picture becomes more exact
when $k_q$ becomes smaller with respect to other constants.

The kinetic equation for the linear chain is
\begin{equation}\label{chainkin}
\dot{c_i}=k_{i-1} c_{i-1}-k_i c_i,
\end{equation}
The coefficient matrix $K$ of these equations is very simple. It
has nonzero elements only on the main diagonal, and one position
below. The eigenvalues of $K$ are $-k_i$ ($i=1,...n-1$) and 0. The
left and right eigenvectors for 0 eigenvalue, $l^0$ and $r^0$,
are:
\begin{equation}\label{chain0eigen}
l^0=(1,1,...1), \;\; r^0=(0,0,...0,1),
\end{equation}
all coordinates of $l^0$ are equal to 1, the only nonzero
coordinate of $r^0$ is $r^0_n$ and we represent vector--column
$r^0$ in row.

The catalytic cycle is one of the most important substructures
that we study in reaction networks. In the reduced form the
catalytic cycle is a set of linear reactions:

$$A_1 \to A_2 \to \ldots A_n \to A_1.$$

Reduced form means that in reality some of these reaction are not
monomolecular and include some other components (not from the list
$A_1, \ldots A_n$). But in the study of the isolated  cycle
dynamics, concentrations of these components are taken as constant
and are included into kinetic constants of the cycle linear
reactions.

For the constant of  elementary reaction $A_i \to $ we use the
simplified notation $k_i$ because the product of this elementary
reaction  is known, it is $A_{i+1}$ for $i<n$ and $A_1$ for $i=n$.
The elementary reaction rate is $w_i = k_i c_i$, where $c_i$ is
the concentration of $A_i$. The kinetic equation is:
\begin{equation}\label{kinCyc}
\dot{c}_i = k_{i-1} c_{i-1} - k_i c_i,
\end{equation}
where by definition $c_0=c_n$, $k_0=k_n$, and $w_0=w_n$. In the
stationary state ($\dot{c}_i = 0$), all the $w_i$ are equal:
$w_i=w$. This common rate $w$ we call the cycle stationary rate,
and
\begin{equation}\label{CycleRate}
w = \frac{b}{\frac{1}{k_1}+\ldots \frac{1}{k_n}}; \; \; c_i
=\frac{w}{k_i},
\end{equation}
where $b=\sum_i c_i$ is the conserved quantity for reactions in
constant volume. Let one of the constants, $k_{\min}$,  be much
smaller than others (let it be $k_{\min} = k_n$):
\begin{equation}\label{LimStepCycle}
k_i \gg k_{\min} \ \ {\rm if} \ \ i\neq n \ .
\end{equation}
In this case, in linear approximation
\begin{equation}\label{CycleLimRateLin}
c_n = b\left(1 -  \sum_{i<n}\frac{k_n}{k_i}\right),  \; c_i = b
\frac{k_n}{k_i},\;  w= k_n b.
\end{equation}

The simplest zero order approximation for the steady state gives
\begin{equation}\label{CycleLimZero}
c_n = b,  \; c_i = 0\; (i \neq n).
\end{equation}
This is trivial: all the concentration is collected at the
starting point of the ``narrow place", but may be useful as an
origin point for various approximation procedures.

So, the stationary rate of a cycle is determined by the smallest
constant, $k_{\min}$, if it is much smaller than the constants of
all other reactions (\ref{LimStepCycle}):
\begin{equation}\label{limitation}
w\approx k_{\min} b .
\end{equation}
In that case we say that the cycle has a limiting step with
constant $k_{\min}$.

There is significant difference between the examples of limiting
steps for the chain of reactions and for irreversible cycle. For
the chain, the steady state does not depend on nonzero rate
constants. It is just $c_n=b, c_1=c_2=...=c_{n-1}=0$. The smallest
rate constant $k_q$ gives the smallest positive eigenvalue, the
relaxation time is $\tau = 1/k_q$. The corresponding approximation
of eigenmode (right eigenvector) $r^1$ has coordinates:
$r^1_1=...=r^1_{q-1}=0$, $r^1_q=1$, $r^1_{q+1}=...=r^1_{n-1}=0$,
$r_n=-1$. This exactly corresponds to the statement that the whole
dynamics in the time scale $\gtrsim 1/k_q$ can be represented by a
single reaction $A_q \to A_n$ with reaction rate constant $k_q$.
The left eigenvector for eigenvalue $k_q$ has approximation $l^1$
with coordinates $l^1_1=l^1_2=...=l^1_q=1$,
$l^1_{q+1}=...=l^1_n=0$. This vector provides the almost exact
{\it lumping} on time scale $\gtrsim 1/k_q$. Let us introduce a
new variable $c_{\rm lump}=\sum_i l_i c_i$, i.e. $c_{\rm lump}
=c_1+c_2+...+c_q$. For the time scale $\gtrsim 1/k_q$ we can write
$c_{\rm lump}+c_n\approx b$, $\D c_{\rm lump} /\D t \approx -k_q
c_{\rm lump}$, $\D c_n /\D t \approx k_q c_{\rm lump}$.

In the example of a cycle, we approximate the steady state, that
is, the right eigenvector $r^0$  for zero eigenvalue (the left
eigenvector is known and corresponds to the main linear balance
$b$: $l^0_i\equiv 1$). In the zero-order approximation, this
eigenvector has coordinates $r^0_1=...=r^0_{n-1}=0$, $r^0_n=1$.

If ${k_n}/{k_i}$ is  small for all $i<n$, then the kinetic
behaviour of the cycle is determined by a linear chain of $n-1$
reactions $A_1 \to A_2 \to ... A_n$, which we obtain after cutting
the limiting step. The characteristic equation for an irreversible
cycle, $\prod_{i=1}^n (\lambda + k_i) - \prod_{i=1}^n k_i =0$,
tends to the characteristic equation for the linear chain,
$\lambda \prod_{i=1}^{n-1} (\lambda + k_i)=0$, when $k_n \to 0$.

The characteristic equation for a cycle with limiting step
($k_n/k_i \ll 1$) has one simple zero eigenvalue that corresponds
to the conservation law $\sum c_i = b$ and $n-1$ nonzero
eigenvalues
\begin{equation}\label{cycle spectra}
\lambda_i = - {k_i} + \delta_i \; (i<n).
\end{equation}
where $\delta_i \to 0$ when $\sum_{i<n}\frac{k_n}{k_i} \to 0$.

A cycle with limiting step (\ref{kinCyc}) has real eigenspectrum
and demonstrates monotonic relaxation without damped oscillations.
Of course, without limitation such oscillations could exist, for
example, when all $k_i\equiv k>0$, ($i=1,...n$).

The relaxation time of a stable linear system (\ref{kinCyc}) is,
by definition, $\tau = 1/\min \{Re (-\lambda_i)\}$ ($\lambda \neq
0$). For small $k_n$, $\tau \approx 1/k_{\tau}$, $k_{\tau}=\min
\{k_i\}$, ($i=1,... n-1$). In other words, for a cycle with
limiting step, $k_{\tau}$ is the {\it second slowest rate
constant}: $k_{\min} \ll k_{\tau}\leq ... $.

\section{Results}

\subsection{Model assumptions}

We consider two models of action of microRNA on protein
translation process proposed in \cite{Nissan2008}: the simplest
linear model, and the non-linear model which explicitly takes into
account recycling of ribosomal subunits and initiation factors.

Both models, of course, represent significant simplifications of
biological reality. First, only a limited subset of all possible
mechanisms of microRNA action on the translation process is
considered (see Fig.~\ref{translation_miRNA}). Second, all
processes of synthesis and degradation of mRNA and protein are
deliberately neglected. Third, interaction of microRNA and mRNA is
simplified: it is supposed that when microRNA is added to the
experimental system then only mRNA with bound microRNAs are
present (this also assumes that the concentration of microRNA is
abundant with respect to mRNA). Concentrations of microRNA and
mRNA are supposed to be constant. Interaction of only one type of
microRNA and one type of mRNA is considered (not a mix of several
microRNAs). The process of initiation is greatly simplified: all
initiation factors are represented by only one molecule which is
marked as eIF4F.

Finally, the classical chemical kinetics approach is applied,
based on solutions of ordinary differential equations, which
supposes sufficient and well-stirred amount of both microRNAs and
mRNAs. Another assumption in the modeling is the mass action law
assumed for the reaction kinetic rates.

It is important to underline the interpretation of certain
chemical species considered in the system. The ribosomal subunits
and the initiation factors in the model exist in free and bound
forms, moreover, ribosomal subunits can be bound to several
regions of mRNA (the initiation site, the start codon, the coding
part). Importantly, several copies of fully assembled ribosome can
be bound to one mRNA. To model this situation, we have to
introduce the following quantification rule for chemical species:
amount of ''ribosome bound to mRNA`` means the total number of
ribosomes translating proteins, which is not equal to the number
of mRNAs with ribosome sitting on them, since one mRNA can hold
several translating ribosomes (polyribosome). In this view, mRNAs
act as {\it places} or {\it catalyzers}, where translation takes
place, whereas mRNA itself formally is not consumed in the process
of translation, but, of course, can be degraded or synthesized
(which is, however, not considered in the models described
further).

\subsection{The simplest linear protein translation model}

\begin{figure}
\centerline{a)\includegraphics[width=11cm,
height=6cm]{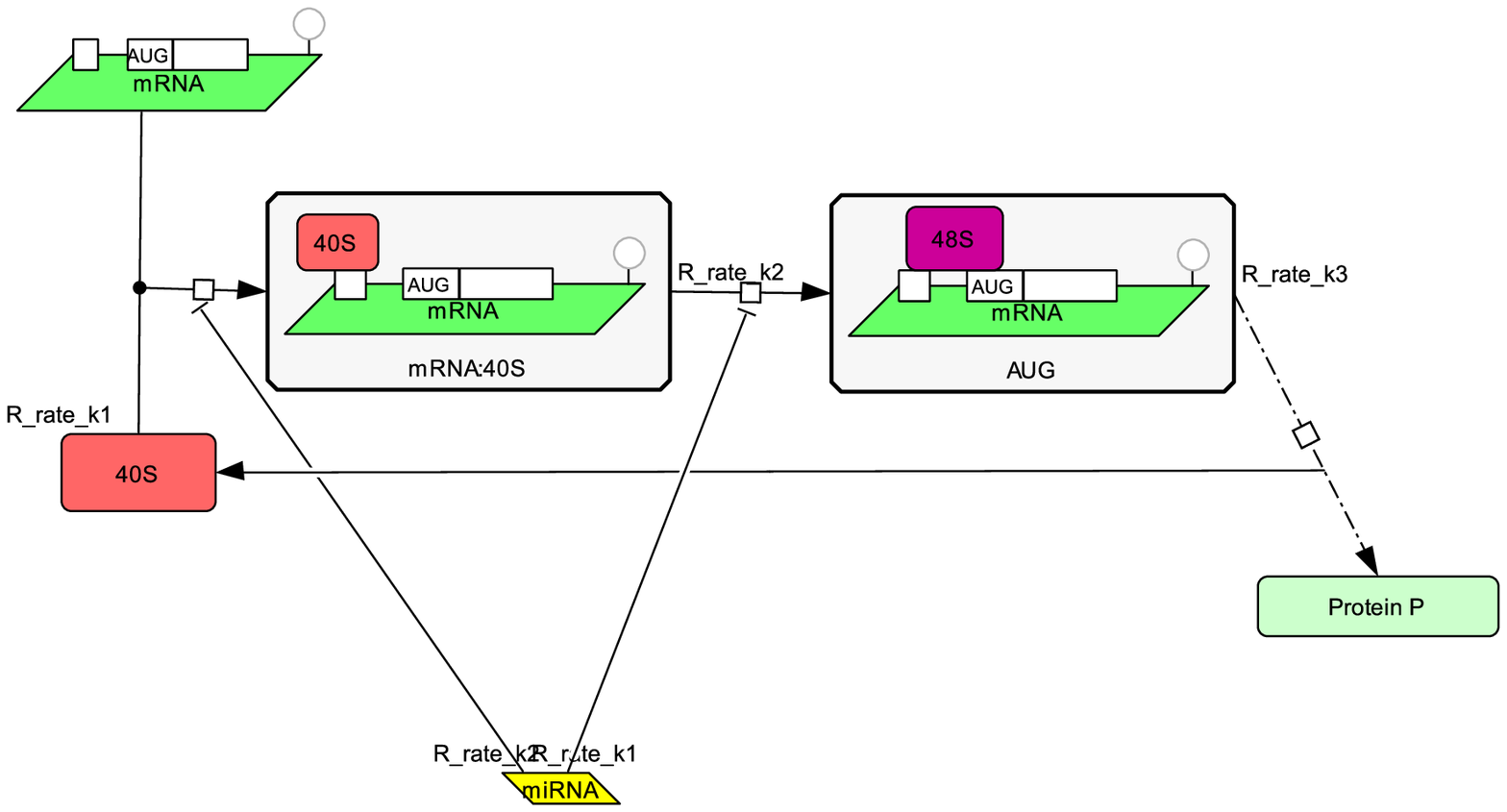}b)\includegraphics[width=5cm,height=4cm]{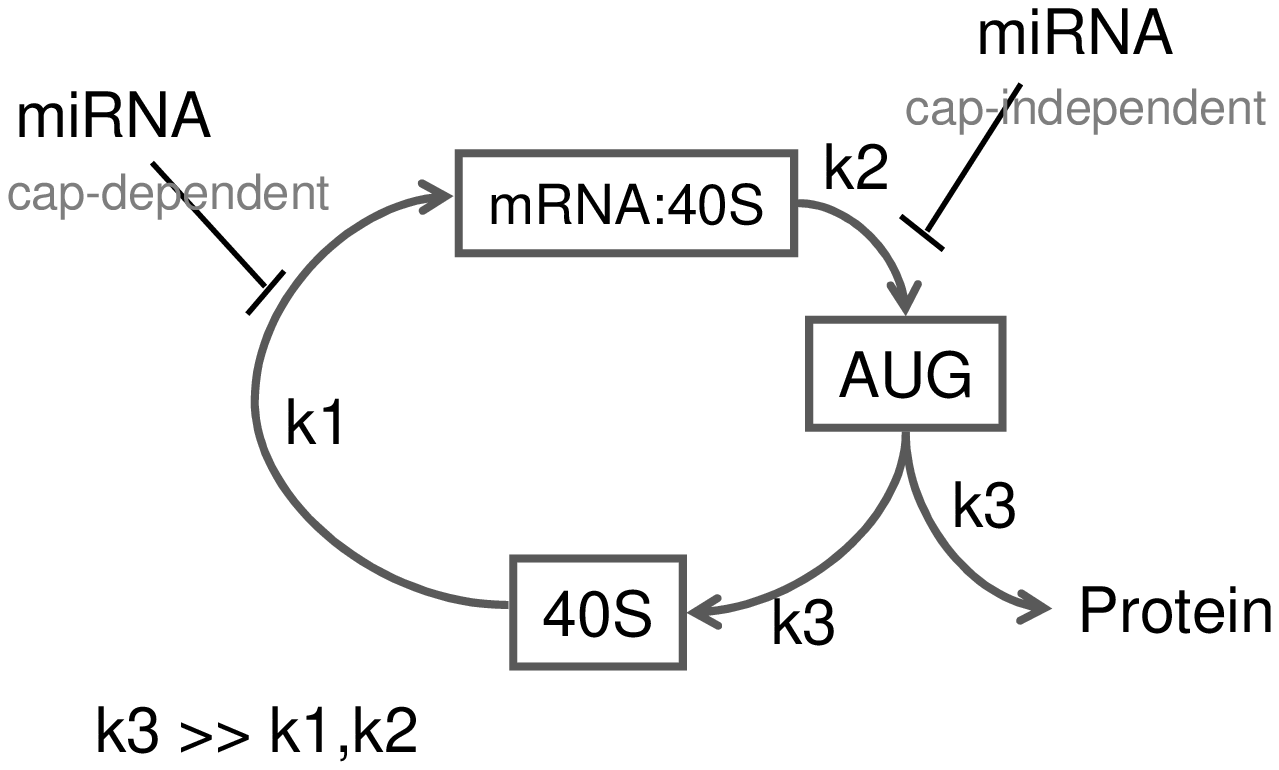}
} \caption{\label{LinearParker} The simplest model of microRNA
action on the protein translation, represented with use of Systems
Biology Graphical Notation (a) and schematically with the
condition on the constants (b). The two mechanism of microRNA
action.}
\end{figure}

The simplest representation of the translation process has the
form of a circular cascade of reactions \cite{Nissan2008} (see
Fig.~\ref{LinearParker}).

The list of chemical species in the model is the following:

1. 40S, free small ribosomal subunit.

2. mRNA:40S, small ribosomal subunit bound to the iniation site.

3. AUG, small ribosomal subunit bound to the start codon.

The catalytic cycle is formed by the following reactions:

1. 40S $\rightarrow$ mRNA:40S, Initiation complex assembly (rate
$k1$).

2. mRNA:40S $\rightarrow$ AUG, Some late and cap-independent
initiation steps, such as scanning the 5'UTR by for the start
codon AUG recognition (rate $k2$).

3. AUG $\rightarrow$ 40S, combined processes of 60S ribosomal unit
joining and protein elongation, which leads to production of the
protein (rate $k3$), and fall off of the ribosome from mRNA.

The model is described by the following system of equations
\cite{Nissan2008}:

\begin{equation}
\left\{
\begin{split}
   &\frac{d\thinspace [40S](t)}{dt} = -k_1[40S]+k_3[AUG] \\
   &\frac{d\thinspace [mRNA:40S](t)}{dt} = k_1[40S]-k_2[mRNA:40S]    \\
   &\frac{d\thinspace [AUG](t)}{dt} = k_2[mRNA:40S]-k_3[AUG] \\
   &Prsynth(t) = k_3[AUG](t)
\end{split}
\right.\label{ParkerLinearEquations}
\end{equation}

\noindent where $Prsynth(t)$ is the rate of the protein synthesis.

Following \cite{Nissan2008}, let us assume that $k_3 \gg k_1,
k_2$. This choice was justified by the following statement:
``...The subunit joining and protein production rate (k3) is
faster than k1 and k2 since mRNA–40S complexes bound to the AUG
without the 60S subunit are generally not observed in translation
initiation unless this step is stalled by experimental methods,
and elongation is generally thought to not be rate limiting in
protein synthesis.." \cite{Nissan2008}.

Under this condition, the equations (\ref{ParkerLinearEquations})
have the following approximate solution (which becomes the more
exact the smaller the $\frac{k_1+k_2}{k_3}$ ratio):

\begin{equation}
\left[
\begin{array}{lllll}
~[40S](t) \\ ~[mRNA:40S](t) \\ ~[AUG](t)
\end{array}
\right] = \frac{[40S]_0}{\frac{1}{k_1}+\frac{1}{k_2}} \left(
\left[
\begin{array}{lllll}
1/k_1 \\ 1/k_2 \\ 1/k_3
\end{array}
\right]+\frac{1}{k_3} \left[
\begin{array}{lllll}
-1 \\ 1 \\ 0
\end{array}
\right]e^{-k_3t}+\frac{1}{k_2}\left[
\begin{array}{lllll}
0 \\ 1 \\ -1
\end{array}
\right]e^{-(k_1+k_2)t}  \right) \label{ParkerLinearSolution}
\end{equation}

\begin{equation}
Prsynth(t) =
\frac{[40S]_0}{\frac{1}{k_1}+\frac{1}{k_2}}\left(1-\frac{k_3}{k_2}e^{-(k_1+k_2)t}\right)
\label{ParkerLinearSolution1}
\end{equation}

\noindent for the initial condition

\begin{equation}
\left[
\begin{split}
&[40S] \\ &[mRNA:40S] \\ &[AUG] \\ &Prsynth
\end{split}
\right]_0 = \left[
\begin{split}
[40S]_0\\ 0\\ 0\\ 0
\end{split}
\right] \label{ParkerLinearInitialCondition} .\end{equation}

From the solution (\ref{ParkerLinearSolution}) it follows that the
dynamics of the system evolves on two time scales: 1) fast
elongation dynamics on the time scale $\approx 1/k_3$; and 2)
relatively slow translation initiation dynamics with the
relaxation time $ t_{rel} \approx \frac{1}{k_1+k_2}$. The protein
synthesis rate formula (\ref{ParkerLinearSolution1}) does not
include the $k_3$ rate, since it is neglected with respect to
$k_1, k_2$ values. From (\ref{ParkerLinearSolution1}) we can
extract the formula for the protein synthesis steady-state rate
$Prsynth$ (multiplier before the parentheses) and the relaxation
time $t_{rel}$ for it (inverse of the exponent power):

\begin{equation}
Prsynth = \frac{[40S]_0}{\frac{1}{k_1}+\frac{1}{k_2}},\hspace{1cm}
t_{rel} = \frac{1}{k_1+k_2}
\end{equation}

Now let us consider two experimental situations: 1) the rates of
the two translation initiation steps are comparable $k_1 \approx
k_2$; 2) the cap-dependent rate $k_1$ is limiting: $k_1 \ll k_2$.
Accordingly to \cite{Nissan2008}, the second situation can
correspond to modified mRNA with an alternative cap-structure,
which is much less efficient for the assembly of the initiation
factors, 40S ribosomal subunit and polyA binding proteins.

\begin{figure}
\centerline{\includegraphics[width=7cm,
height=5cm]{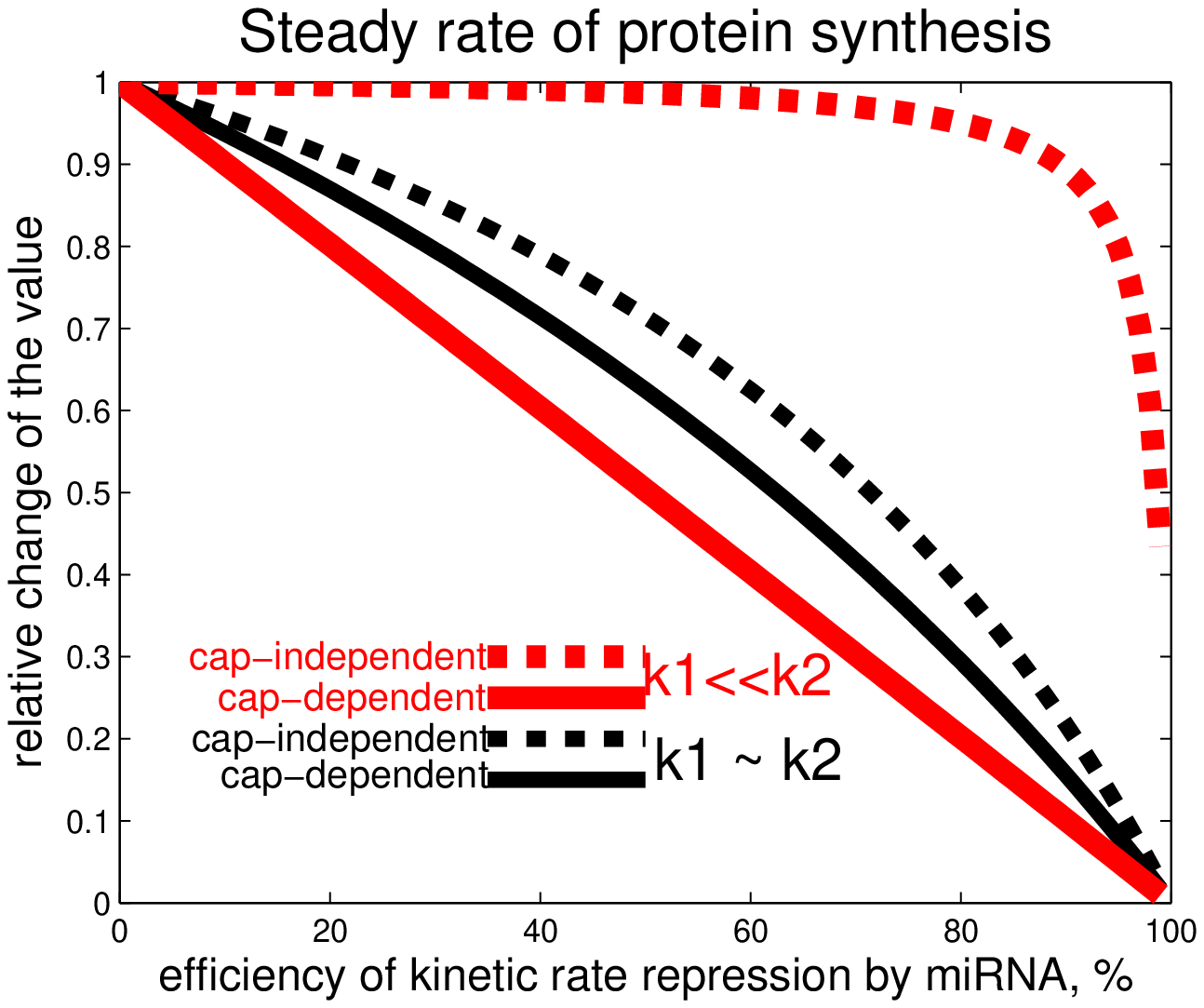}\includegraphics[width=7cm,
height=5cm]{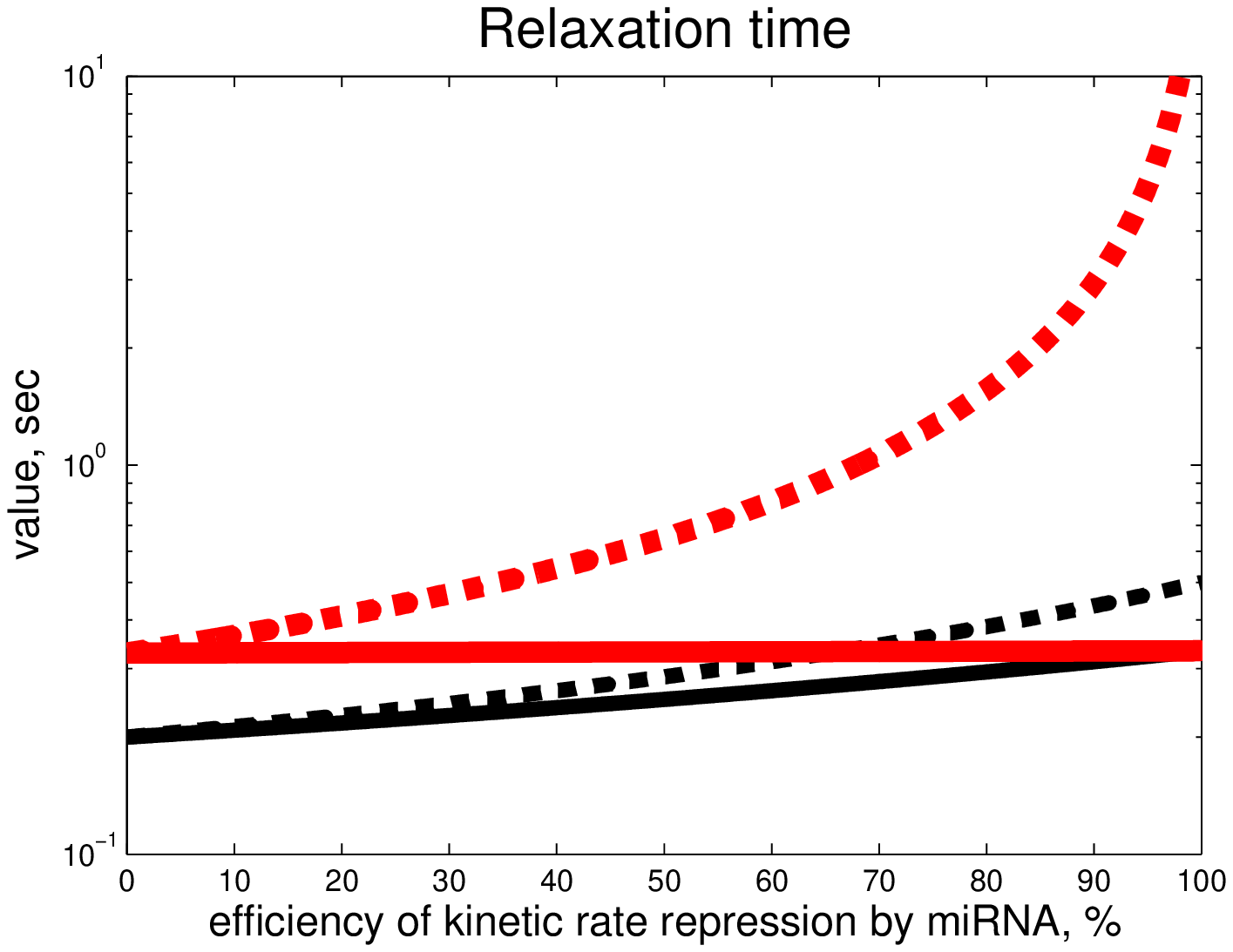}}
\caption{\label{LinearParkerInhibition} Graphs illustrating the
predicted change in the steady-state rate of protein synthesis
(left), and its relaxation time (i.e., the time needed to recover
from a perturbation to the steady state value), right. Four curves
are presented. The black ones are for the wild-type cap structure,
which is modeled by $k_1=k_2$. The red ones are for the modified
structure, when $k_1 \ll k_2$. The main conclusion from the left
graph is that if microRNA acts on a late initiation step,
diminishing $k_2$ then it's effect is not measurable unless $k_2$
is very strongly suppressed (as reported in \cite{Nissan2008}).
The main conclusion from the right is that the effect of microRNA
should be measurable in this case if one looks at dynamical
features such the relaxation time. }
\end{figure}

For these two experimental systems (let us call them 'wild-type'
and 'modified' correspondingly), let us study the effect of
microRNA action. We will model the microRNA action as a mechanism
which effectively diminish the value of a kinetic rate
coefficient. Let us assume that there are two alternative
mechanisms: 1) microRNA acts in a cap-dependent manner (thus,
reducing the $k_1$ constant) and 2) microRNA acts in a
cap-independent manner, for example, through interfering with 60S
subunit joining (thus, reducing the $k_2$ constant). The
dependence of the steady rate of protein synthesis $Prsynth \sim
\frac{1}{\frac{1}{k1}+\frac{1}{k2}}$ and the relaxation time $
t_{rel} \approx \frac{1}{k_1+k_2}$ on the efficiency of the
microRNA action (i.e., how much it is capable to diminish a rate
coefficient) is shown on Fig.~\ref{LinearParkerInhibition}.

Interestingly, experiments with cap structure replacement were
made and the effect of microRNA action on the translation was
measured \cite{Mathonnet2007,Thermann2007}. No change in the
protein rate synthesis after applying microRNA was observed. From
this it was concluded that microRNA in this system should act
through cap-dependent mechanism (i.e., the normal 'wild-type' cap
is required for microRNA recruitement). It was argued that this
could be a misinterpretation \cite{Nissan2008} since in the
'modified' system, cap-dependent translation initiation is a rate
limiting process ($k_1 \ll k_2$), hence, even if microRNA acts in
the cap-independent manner (inhibiting $k_2$), it will have no
effect on the final steady state protein synthesis rate. They
confirmed this by the graph similar to the
Fig.~\ref{LinearParkerInhibition}a.

From the analytical solution (\ref{ParkerLinearSolution}) we can
further develop this idea and claim that it is still possible to
detect the action of microRNA in the 'modified' system if one
measures the protein synthesis relaxation time: if it
significantly increases then microRNA probably acts in the
cap-independent manner despite the fact that the steady state rate
of the protein synthesis does not change (see the
Fig.~\ref{LinearParkerInhibition}b). This is a simple consequence
of the fact that the relaxation time in a cycle of biochemical
reactions is limited by the second slowest reaction
\cite{Gorban2009}. If the relaxation time is not changed in the
presence of microRNA then we can conclude that none of the two
alternative mechanisms of microRNA-based translation repression is
activated in the system, hence, microRNA action is dependent on
the structure of the 'wild-type' transcript cap.

The observations from the Fig.~\ref{LinearParkerInhibition} are
recapitulated in the Table~\ref{tableLinear}. This analysis (of
course, over-simplified in many aspects) provides us with an
important lesson: observed dynamical features of the translation
process with and without presence of microRNA can give clues on
the mechanisms of microRNA action and help to distinguish them in
a particular experimental situation. Theoretical analysis of the
translation dynamics highlights what are the important
characteristics of the dynamics which should be measured in order
to infer the possible microRNA mechanism.

This conclusion suggests the notion of a {\it kinetic signature of
microRNA action mechanism} which we define as {\it the set of
measurable characteritics of the translational machinery dynamics
(features of time series for protein, mRNA, ribosomal subunits
concentrations) and the predicted tendencies of their changes as a
response to microRNA action through a particular biochemical
mechanism}.

\begin{table}
\caption{Modeling of two mechanisms of microRNA action in the
simplest linear model. microRNA action effect is described for the
protein synthesis steady rate and the relaxation time. It is
assumed that the ribosome assembly+elongation step in protein
translation, described by the $k_3$ rate constant, is not rate
limiting \label{tableLinear}} \centering{
\begin{tabular}{|c|c|c|c|}\hline
{Observable value} & {Initiation($k_1$)} & {Step after
initiation,cap-independent($k_2$)} & {Elongation ($k_3$)}
\\\hline \multicolumn{4}{c}{{\bf Wild-type cap}}
\\ {\it Steady-state rate} & decreases & decreases
& no change \\ {\it Relaxation time} & increases slightly &
increases slightly & no change
\\
\multicolumn{4}{c}{{\bf A-cap}} \\ {\it Steady-state rate} &
decreases & no change & no change
\\ {\it Relaxation time} & no change & increases drastically & no
change\\ \hline
\end{tabular}}
\end{table}

\subsection{The non-linear protein translation model}

To explain the effect of microRNA interference with translation
initiation factors, a non-linear version of the translation model
was proposed \cite{Nissan2008} which explicitly takes into account
recycling of initiation factors (eIF4F) and ribosomal subunits
(40S and 60S).

\begin{figure}
\centerline{a)\includegraphics[width=9cm,
height=4cm]{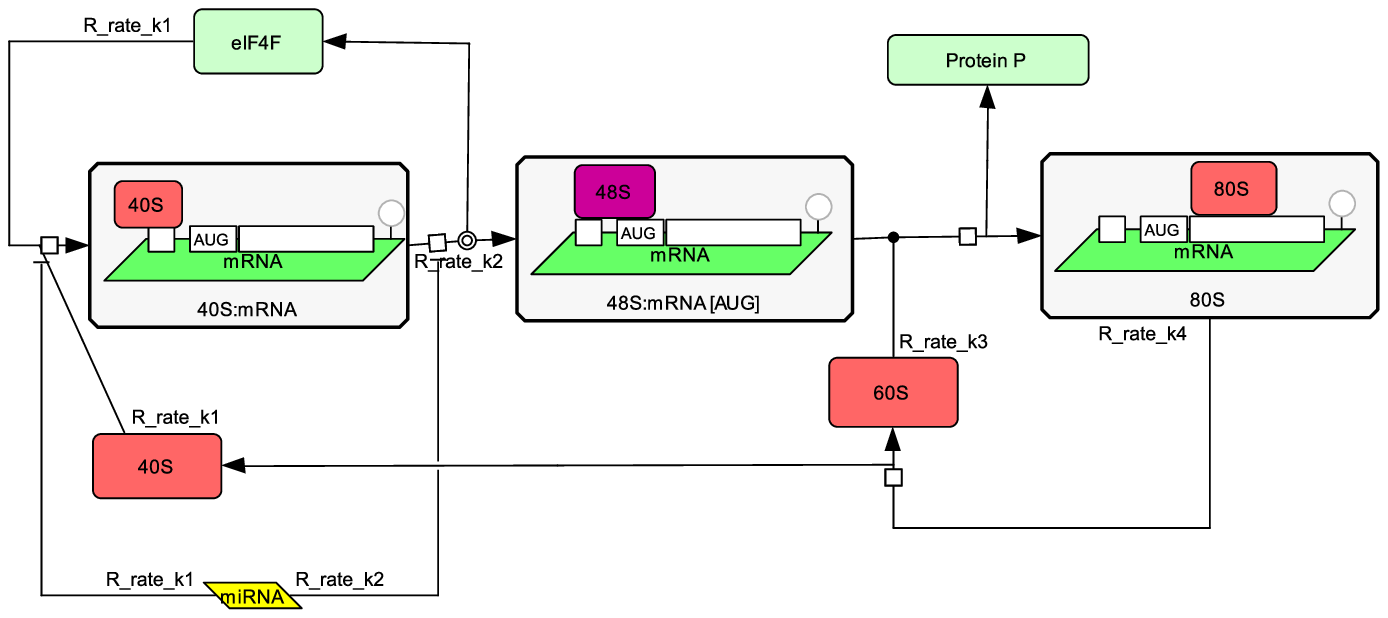}b)\includegraphics[width=5cm,height=4.5cm]{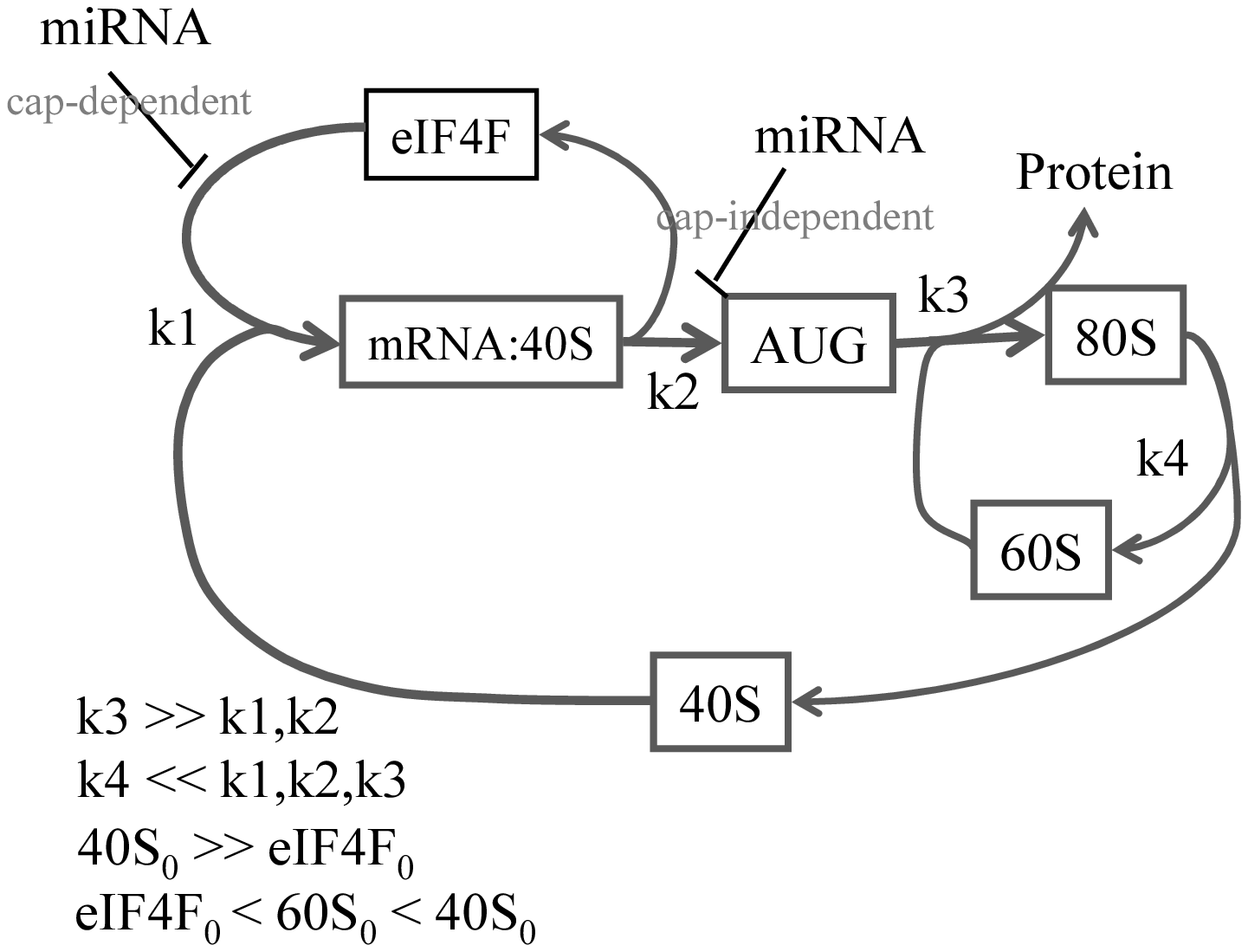}
} \caption{\label{NonLinearParker} Non-linear model of microRNA
action on the protein translation, represented with use of Systems
Biology Graphical Notation (a) and schematically with the
condition on the constants (b). The difference from the simplest
model (\ref{LinearParker}) is in the explicit description of
initiation factors eIF4F, and ribosomal subunits 40S and 60S
recycling.}
\end{figure}

The model contains the following list of chemical species (see
also Fig.~\ref{NonLinearParker}):

1. 40S, free 40S ribosomal subunt.

2. 60S, free 60S ribosomal subunit.

3. eIF4F, free initiation factor.

4. mRNA:40S, formed initiation complex (containing 40S and the
initiation factors), bound to the initiation site of mRNA.

5. AUG, initiation complex bound to the start codon of mRNA.

6. 80S, fully assembled ribosome translating protein.

There are four reactions in the model, all considered to be
irreversible:

1. 40S + eIF4F $\rightarrow$ mRNA:40S, assembly of the initiation
complex (rate $k_1$).

2. mRNA:40S $\rightarrow$ AUG, some late and cap-independent
initiation steps, such as scanning the 5'UTR by for the start
codon AUG recognition (rate $k_2$).

3. AUG $\rightarrow$ 80S, assembly of ribosomes and protein
translation (rate $k_3$).

4. 80S $\rightarrow$ 60S+40S, recycling of ribosomal subunits
(rate $k_4$).

The model is described by the following system of equations
\cite{Nissan2008} :

\begin{equation}
\left\{ \begin{split}
   &\frac{d\thinspace [40S]}{dt} = -k_1[40S]~[eIF4F]+k_4[80S] \\
   &\frac{d\thinspace [eIF4F]}{dt} = -k_1[40S]~[eIF4F]+k_2[mRNA:40S] \\
   &\frac{d\thinspace [mRNA:40S]}{dt} = k_1[40S]~[eIF4F]-k_2[mRNA:40S]    \\
   &\frac{d\thinspace [AUG]}{dt} = k_2[mRNA:40S]-k_3[AUG]~[60S] \\
   &\frac{d\thinspace [60S]}{dt} = -k_3[AUG]~[60S]+k_4[80S]\\
   &\frac{d\thinspace [80S]}{dt} = k_3[AUG]~[60S]-k_4[80S]\\
   &Prsynth(t) = k_3[AUG]~[60S]
   \end{split}
\right. \label{ParkerNonlinearEquations}
\end{equation}

\noindent where $[40S]$ and $[60S]$ are the concentrations of free
40S and 60S ribosomal subunits, $[eIF4F]$ is a concentration of
free translation initiation factors, $[mRNA:40S]$ is the
concentration of 40S subunit bound to the initiation site of mRNA,
$[AUG]$ is the concentration of the initiation complex bound to
the start codon, $[80S]$ is the concentration of ribosomes
translating protein, and $Prsynth$ is the rate of protein
synthesis.

The model (\ref{ParkerNonlinearEquations}) contains three independent conservations laws:

\begin{equation}
   [mRNA:40S] + [40S] + [AUG] + [80S] = [40S]_0
 \label{ParkerNonlinearConservationLaw40S},
\end{equation}

\begin{equation}
   [mRNA:40S] + [eIF4F] = [eIF4F]_0
\label{ParkerNonlinearConservationLawEIF},
\end{equation}

\begin{equation}
   [60S] + [80S] = [60S]_0
\label{ParkerNonlinearConservationLaw60S},
\end{equation}

The following assumptions on the model parameters were suggested
in \cite{Nissan2008}:

\begin{equation}
k_4\ll k_1,k_2,k_3; k_3\gg k_1,k_2; [eIF4F]_0\ll [40S]_0;
[eIF4F]_0 < [60S]_0 < [40S]_0
\label{ParkerNonlinearParameterOrders}
\end{equation}

\noindent with the following justification: ''...The amount 40S
ribosomal subunit was set arbitrarily high ... as it is thought to
generally not be a limiting factor for translation initiation. In
contrast, the level of eIF4F, as the canonical limiting factor,
was set significantly lower so translation would be dependent on
its concentration as observed experimentally... Finally, the
amount of subunit joining factors for the 60S large ribosomal
subunit were estimated to be more abundant than eIF4F but still
substoichiometric when compared to 40S levels, consistent with in
vivo levels... The k4 rate is relatively slower than the other
rates in the model; nevertheless, the simulation's overall protein
production was not altered by changes of several orders of
magnitude around its value... `` \cite{Nissan2008}.

Notice that further in our paper we show that the last statement
about the value of $k_4$ is needed to be made more precise: in the
model by Nissan and Parker, $k_4$ {\it is a critical parameter}.
It does not affect the steady state protein synthesis rate only in
one of the possible scenarios (inefficient initiation, deficit of
the initiation factors).

\subsubsection{Steady state solution}

The final steady state of the system can be calculated from the
conservation laws and the balance equations among all the reaction
fluxes:

\begin{equation}
    k_2\cdot [mRNA:40S]_s = k_3\cdot [AUG]_s\cdot [60S]_s = k_4\cdot
    [80S]_s = k_1\cdot [40S]_s\cdot [eIF4F]_s
\label{ParkerNonlinearSteadyBalance},
\end{equation}

\noindent where 's' index stands for the steady state value. Let
us designate a fraction of the free [60S] ribosomal subunit in the
steady state as $x=\frac{[60S]_s}{[60S]_0}$. Then we have

\begin{equation}
  \begin{split}
  &[mRNA:40S]_s = \frac{k_4}{k_2}[60S]_0(1-x), [AUG]_s =
  \frac{k_4}{k_3}\left(\frac{1-x}{x}\right), \\ &[eIF4F]_s = [eIF4F]_0-\frac{k_4}{k_2}[60S]_0(1-x), \\
  &[60S]_s=[60S]_0x, [80S]_s = [60S]_0(1-x), \\
  &[40S]_s = [40S]_0-[60S]_0(1-x)(1+\frac{k_4}{k_2})-\frac{k_4}{k_3}\left(\frac{1-x}{x}\right)
  \end{split}
\label{ParkerNonlinearSteadyState}
\end{equation}

\noindent and the equation to determine $x$, in which we have
neglected the terms of smaller order of magnitude, based on conditions (\ref{ParkerNonlinearParameterOrders}):

\begin{equation}
  \begin{split}
&
x^3+x^2\left(\alpha+(\delta-1)+(\beta-1)\right)+x\left(-\alpha+(\delta-1)(\beta-1)\right)+\gamma(1-\beta)=0,
\\
&\alpha = \frac{k_2}{k_1[60S]_0},
\beta=\frac{[eIF4F]_0k_2}{[60S]_0k_4}, \gamma =
\frac{k_4}{k_3[60S]_0}, \delta = \frac{[40S]_0}{[60S]_0}
  \end{split}
\label{ParkerNonlinearSteadyStateX}.
\end{equation}

From the inequalities on the parameters of the model, we have
$\delta>1$, $\gamma\ll 1$ and, if $k_1\gg k_4/[eIF4F]_0$ then
$\alpha\ll \beta$. From these remarks it follows that the constant
term $\gamma(1-\beta)$ of the equation
(\ref{ParkerNonlinearSteadyStateX}) should be much smaller than
the other polynomial coefficients, and the equation
(\ref{ParkerNonlinearSteadyStateX}) should have one solution close
to zero and two others:

\begin{equation}
  \begin{split}
&x_0\approx \frac{k4}{k3\cdot\left([40S]_0-[60S]_0\right)},
\\ &x_1\approx 1-\beta+\frac{\alpha\beta}{\delta-\beta} = 1-\frac{[eIF4F]_0k_2}{[60S]_0k_4}+\frac{k_2^2[eIF4F]_0}{k_1k_4[40S]_0}\frac{1}{1-\frac{[eIF4F]_0k_2}{[40S]_0k_4}},
\\
&x_2\approx 1-\frac{[40S]_0}{[60S]_0},
  \end{split}
\label{ParkerNonlinearSteadyStateXs}
\end{equation}

\noindent provided that $\alpha\ll |1-\delta|$ or $\alpha\ll |1-\beta|$. In the expression for $x_1$ we can not neglect
the term proportional to $\alpha$, to avoid zero values in (\ref{ParkerNonlinearSteadyStateX}).

The solution $x_2$ is always negative, which means that one can
have one positive solution $x_0\ll 1$ if
$\frac{[eIF4F]_0k_2}{[60S]_0k_4}\gtrapprox 1$ and two positive
solutions $x_0$ and $x_1$ if
$\frac{[eIF4F]_0k_2}{[60S]_0k_4}\lessapprox 1$. However, from
(\ref{ParkerNonlinearSteadyState}),
(\ref{ParkerNonlinearSteadyStateXs}) and
(\ref{ParkerNonlinearParameterOrders}) it is easy to check that if
$x_1>0$ then $x_0$ does not correspond to positive value of
$[eIF4F]_s$. This means that for a given combination of parameters
satisfying (\ref{ParkerNonlinearParameterOrders}) we can have only
one steady state (either $x_0$ or $x_1$).

The two values $x=x_0$ and $x=x_1$ correspond to two different
modes of translation. When, for example, the amount of the
initiation factors $[eIF4F]_0$ is not sufficient to provide
efficient initiation ($[eIF4F]_0<\frac{k_2}{[60S]_0k_4}$, $x=x_1$)
then most of the $40S$ and $[60S]$ subunits remain in the free
form, the initiation factor $[eIF4F]$ being always the limiting
factor. If the initiation is efficient enough
($[eIF4F]_0>\frac{k_2}{[60S]_0k_4}$), then we have $x=x_0\ll 1$
when almost all $60S$ ribosomal subunits are engaged in the
protein elongation, and $[eIF4F]$ being a limiting factor at the
early stage, however, is liberated after and ribosomal subunits
recycling becomes limiting in the initiation (see the next section
for the analysis of the dynamics).

Let us notice that the steady state protein synthesis rate under
these assumptions is

\begin{equation}
Prsynth = k_4\cdot [60S]_0(1-x)\approx \left\{
\begin{split}
&k_4\cdot [60S]_0,\ if\ \frac{[eIF4F]_0k_2}{[60S]_0k_4}> 1
\\ &k_2\cdot [eIF4F]_0, \ else
\end{split} \right. \label{ParkerNonLinearProtSynthesis}.
\end{equation}

This explains the numerical results obtained in \cite{Nissan2008}:
with low concentrations of $[eIF4F]_0$ microRNA action would be
efficient only if it affects $k_2$ or if it competes with $eIF4F$
for binding to the mRNA cap structure (thus, effectively further
reducing the level $[eIF4F]_0$). With higher concentrations
$[eIF4F]_0$, other limiting factors become dominant: $[60S]_0$
(availability of the heavy ribosomal subunit) and $k_4$ (speed of
ribosomal subunits recycling which is the slowest reaction rate in
the system). Interestingly, in any situation the protein
translation rate does not depend on the value of $k_1$ directly
(of course, unless it does not become ''globally`` rate limiting),
but only through competing with $eIF4F$ (which makes the
difference with the simplest linear protein translation model).

Equation (\ref{ParkerNonLinearProtSynthesis}) explains also some
experimental results reported in \cite{Mathonnet2007}: increasing
the concentration of $[eIF4F]$ translation initiation factor
enhances protein synthesis but its effect is abruptly saturated
above a certain level.

It would be interesting to make some conclusions on the shift of
polysomal profile from the steady state solutions
\ref{ParkerNonlinearSteadyStateXs}. In this model, the number of
ribosomes sitting on mRNA $N_{polysome}$ is defined by
$N_{polysome}=\frac{[80S]}{[mRNA]}$, where $[mRNA]$ is the
concentration of mRNA. However, $[mRNA]$ is not an explicit
dynamical variable in the model, it is implicitly included in
other model constants, such as $k_1$, together with the effective
volume of cytoplasmic space considered in the model. Nevertheless,
the model can predict the relative shift of the polysome profile.
In the steady state

\begin{equation}
N_{polysome} \sim 1-x =  \left\{
\begin{split}
&k_4\cdot [60S]_0,\ if\ \frac{[eIF4F]_0k_2}{[60S]_0k_4}> 1
\\ &k_2\cdot [eIF4F]_0, \ else
\end{split} \right. \label{ParkerNonLinearPolysome}.
\end{equation}

\noindent and $N_{polysome}$ changes in the same way as the
protein synthesis steady state value.

\subsubsection{Analysis of the dynamics}

It was proposed to use the following model parameters in
\cite{Nissan2008}: $k_1=k_2=2$, $k_3=5$, $k_4=1$, $[40S]_0=100$,
$[60S]_0=25$, $[eIF4F]_0=6$. Although all estimations made in the
previous section remain valid for them, in this section we have
modified the parameter values (anyway very arbitrary estimated) to
make the dynamical features of the model more illustrative. After
we have verified that the semi-analytical approach remains valid
for the set of ''normal`` parameter values from \cite{Nissan2008}
(and, in general, for a very large set of parameters). First, we
modified $k_2=3$ to avoid possible artifact symmetries related to
equal parameter values. Second, we have changed $k_4=0.1$ to have
the case of $x=x_0$ steady state (which, as it will be shown,
corresponds to the three-stage dynamics, which allows to
demonstrate three dominant dynamical systems instead of only one
in the case of $x=x_1$ when $eI4F4$ is a limiting factor during
the whole relaxation process) and $k_3=50$ to have better
separation between $k_3$ and $k_1$, $k_2$.

\begin{figure}
\centerline{a)\includegraphics[width=9cm,
height=7cm]{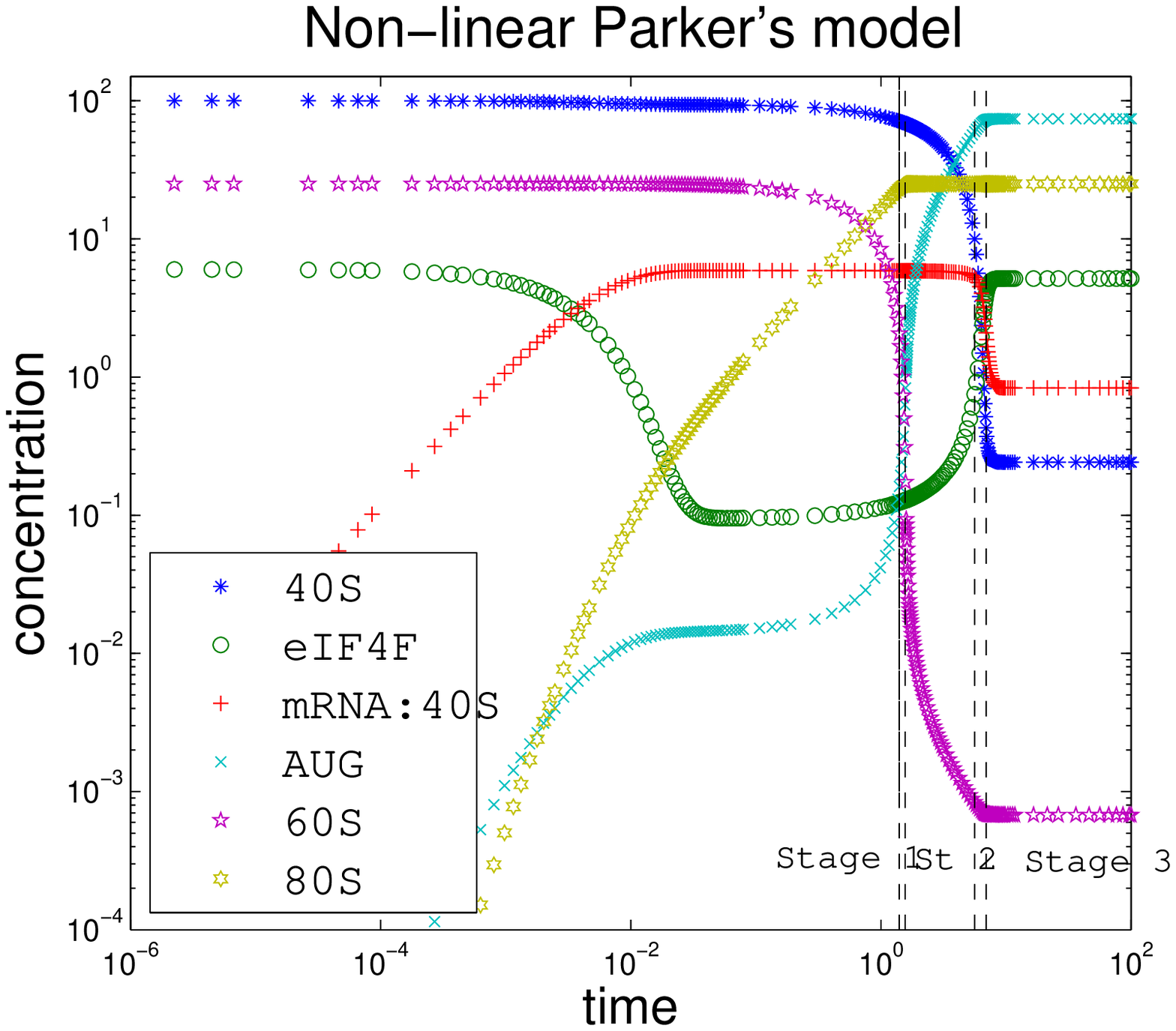}b)\includegraphics[width=9cm,
height=7cm]{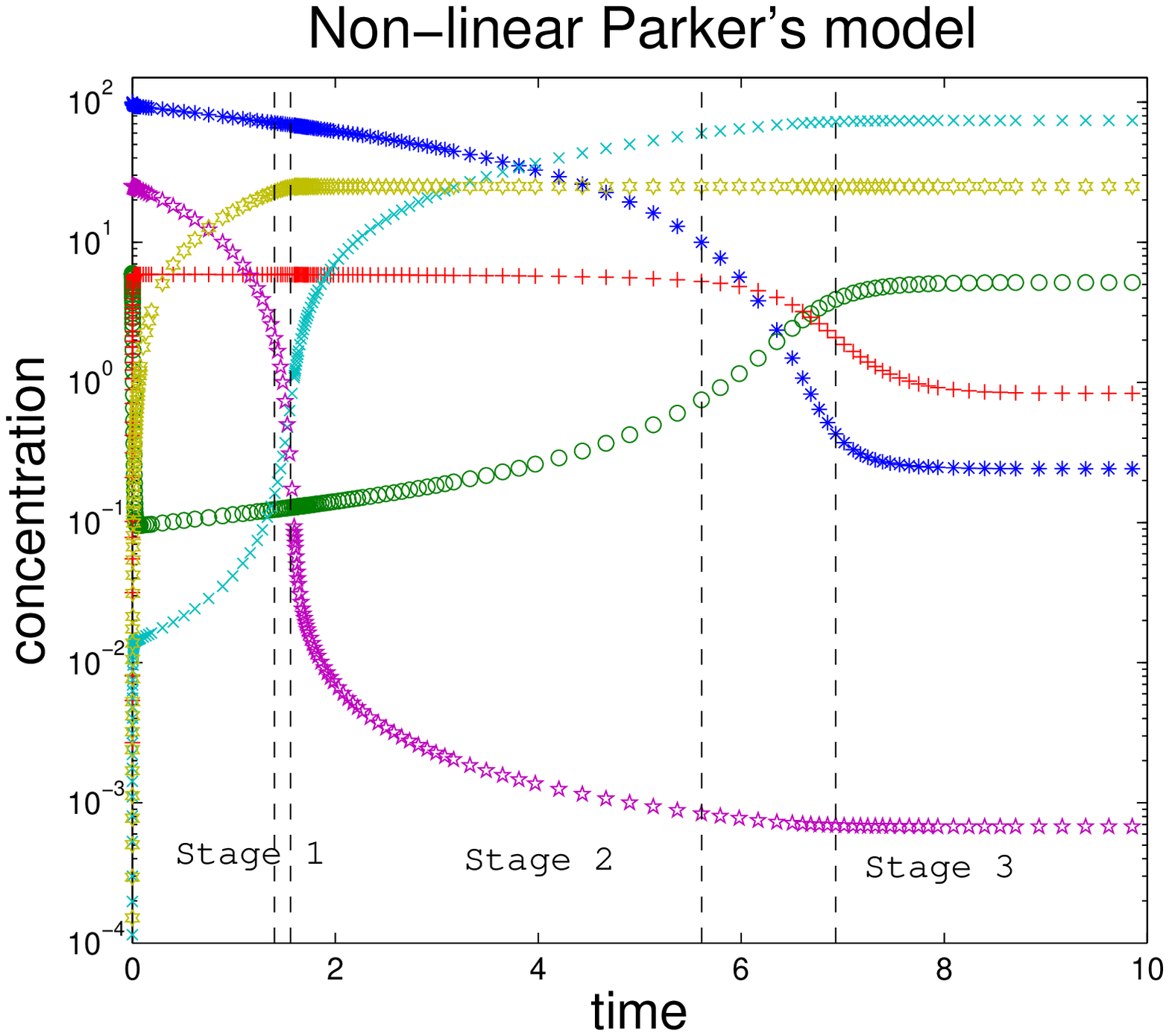}} \centerline{
c)\includegraphics[width=9cm,
height=7cm]{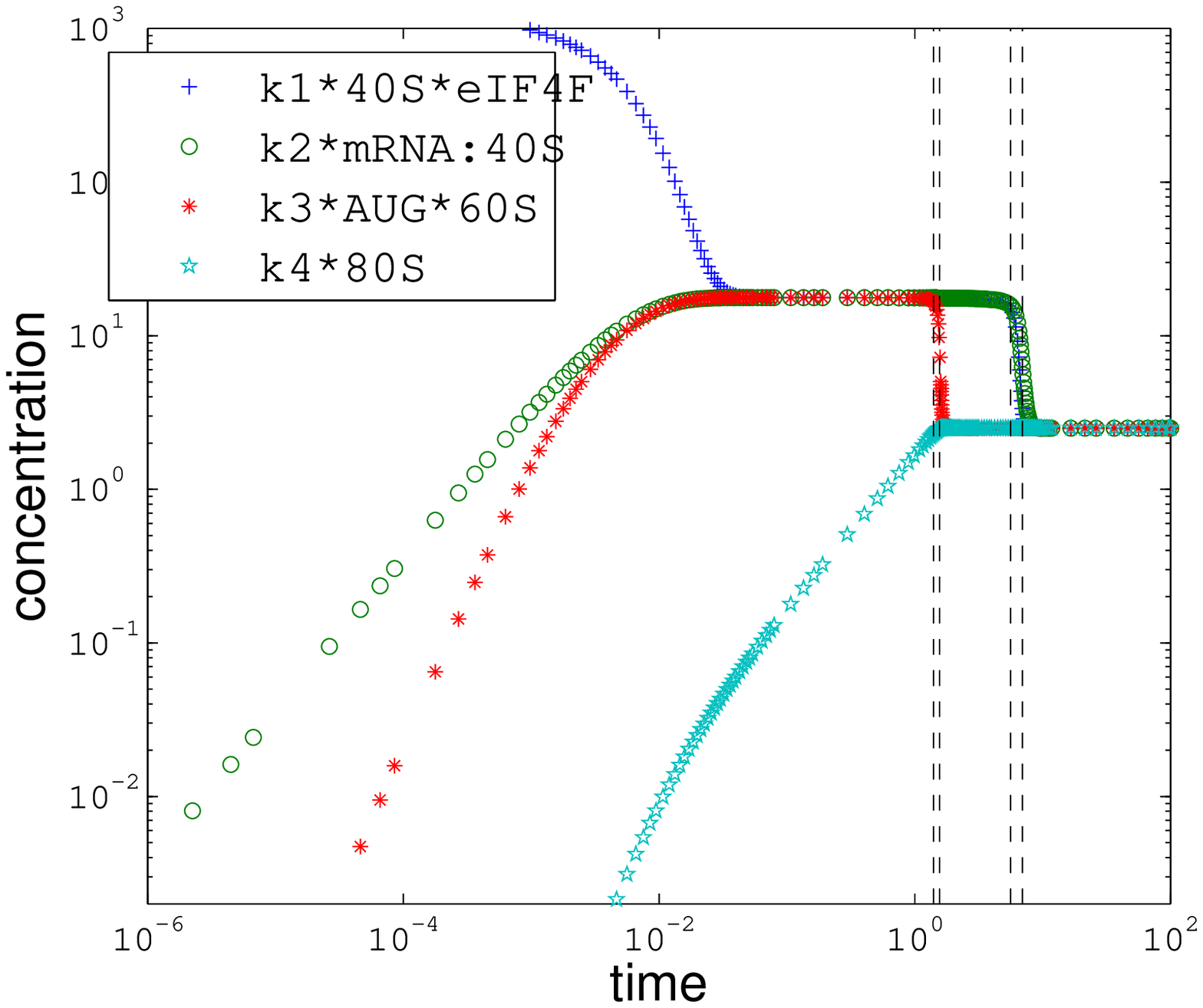}d)\includegraphics[width=9cm,
height=7cm]{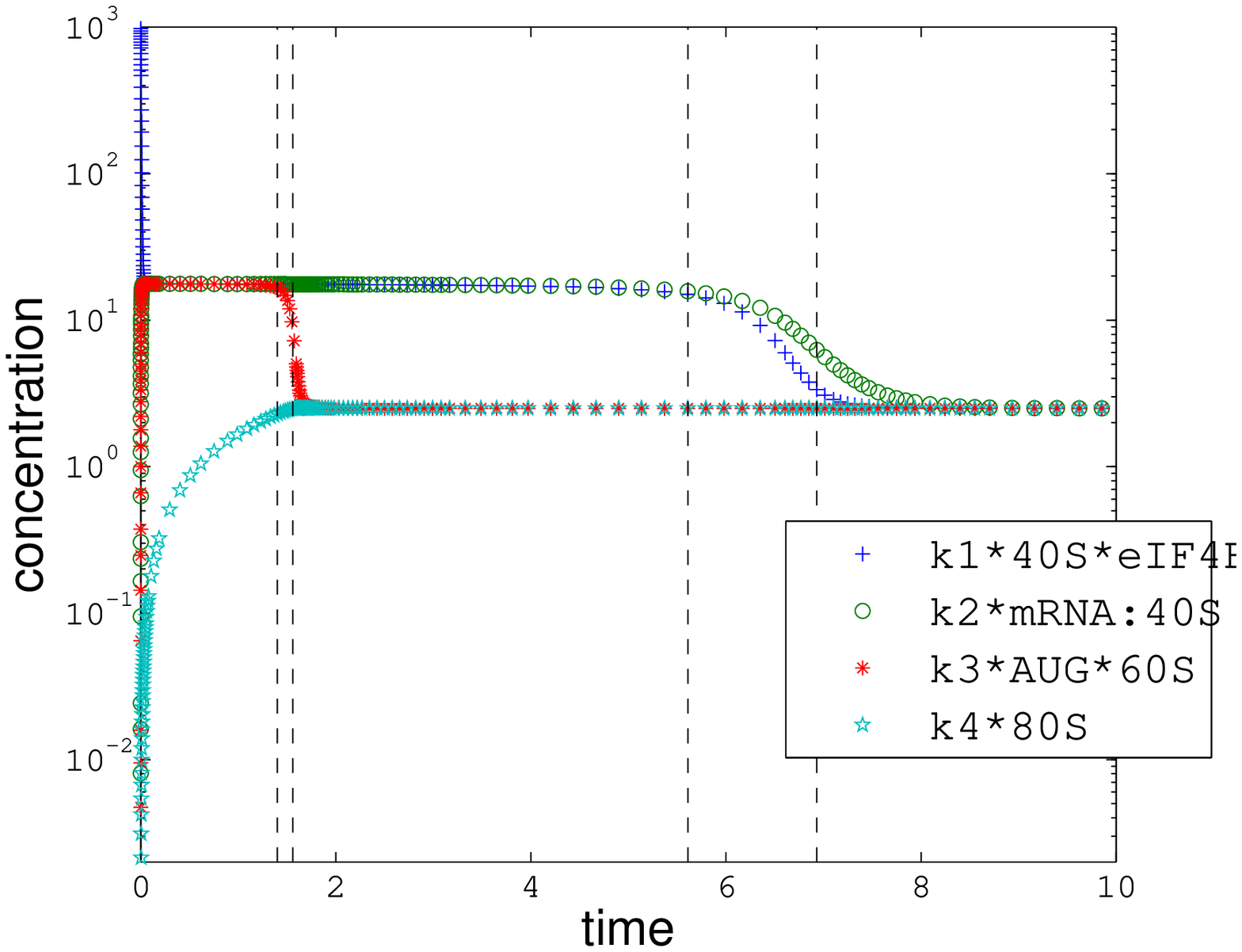}}
\caption{\label{Parker2Solution} Simulation of the non-linear
protein translation model with parameters $k_1=2$, $k_2=3$,
$k_3=50$, $k_4=0.1$, $[40S]_0=100$, $[60S]_0=25$, $[eIF4F]_0=6$.
a) and b) chemical species concentrations at logarithmic and
linear scales; c) and d) reaction fluxes at logarithmic and linear
scales. By the dashed line several stages are delimited during
which the dynamics can be considered as (pseudo-)linear. To
determine where ''$\gg$`` and ''$\ll$`` conditions are violated,
we arbitrarily consider ''much bigger`` or ''much smaller`` as
difference in one order of magnitude (by factor 10).}
\end{figure}

Simulations of the protein translation model with these parameters and the
initial conditions

\begin{equation}
\left[
\begin{split}
&[40S]\\ &[eIF4F] \\ &[mRNA:40S] \\ &[AUG] \\ &[80S] \\ &[60S]
\end{split}
\right]_0 = \left[
\begin{split}
&[40S]_0 \\ &[eIF4F]_0 \\ 0\\ 0\\ 0 \\ &[60S]_0
\end{split}
\right] \label{ParkerNonLinearInitialCondition} .\end{equation}

\noindent are shown on the Fig.~\ref{Parker2Solution}. The
system shows non-trivial relaxation process which takes place in
several epochs. One can check that the ''efficient translation initiation`` ($x=x_0\ll 1$) scenario is realized in this case.
Qualitatively we can distinguish the following stages:

1) Stage 1: Relatively fast relaxation with conditions $[40S]\gg
[eIF4F]$, $[60S]\gg [AUG]$. During this stage, the two non-linear
reactions $40S+eIF4F\rightarrow mRNA:40S$ and $AUG+60S\rightarrow
80S$ can be considered as pseudo-monomolecular ones:
$eIF4F\rightarrow mRNA:40S$ and $AUG\rightarrow 80S$ with rate
constants dependent on $[40S]$ and $[60S]$ respectively. This
stage is characterized by rapidly establishing quasiequilibrium of
three first reactions (R1, R2 and R3 with $k_1$, $k_2$ and $k_3$
constants). Biologically, this stage corresponds to assembling of
the translation initiation machinery, scanning for the start codon
and assembly of the first full ribosome at the start codon
position.

2) Transition between Stage 1 and Stage 2.

3) Stage 2: Relaxation with the conditions $[40S]\gg [eIF4F]$,
$[60S]\ll [AUG]$. During this stage, the reactions
$40S+eIF4F\rightarrow mRNA:40S$ and $AUG+60S\rightarrow 80S$ can
be considered as pseudo-monomolecular $eIF4F\rightarrow mRNA:40S$
and $60S\rightarrow 80S$. This stage is characterized by two local
quasi-steady states established in the two network reaction cycles
(formed from R1-R2 and R3-R4 reactions). Biologically, this stage
corresponds to the first round of elongation, when first ribosomes
moves along the coding region of mRNA. The small ribosomal subunit
40S is still in excess which keeps the initiation stage (reaction
R1-R2 fluxes) relatively fast.

4) Transition between Stage 2 and Stage 3.

5) Stage 3: Relaxation with the conditions $[40S]\ll [eIF4F]$,
$[60S]\ll [AUG]$. During this stage, the reactions
$40S+eIF4F\rightarrow mRNA:40S$ and $AUG+60S\rightarrow 80S$ can
be considered as pseudo-monomolecular $40S\rightarrow mRNA:40S$
and $60S\rightarrow 80S$. During this stage all reaction fluxes
are balanced. Biologically, this stage corresponds to the stable
production of the protein with constant recycling of the ribosomal
subunits. Most of ribosomal subunits 40S are involved in protein
elongation, so the initiation process should wait until the end of
elongation for that they would be recycled.

Stages 1-3 can be associated with the corresponding dominant
systems \cite{Gorban2009} which are shown on
Fig.~\ref{Parker2DominantSubsystems}.

Below we give a more detailed analysis of stages 1-3 and transitions between them.

\subsubsection{Stage 1: translation initiation and assembly of the
first ribosome at the start codon}

The dominant system of the Stage 1
(Fig.~\ref{Parker2DominantSubsystems}a) can be modeled as a linear
system of equations (notice that it is not equivalent to the
system of equations that would correspond to fully monomolecular
reaction network because the reaction R2 is still bimolecular
despite the fact that the products of this reaction do not
interact, which leads to the linear description):

\begin{figure}
\centerline{
\includegraphics[width=5cm,height=3cm]{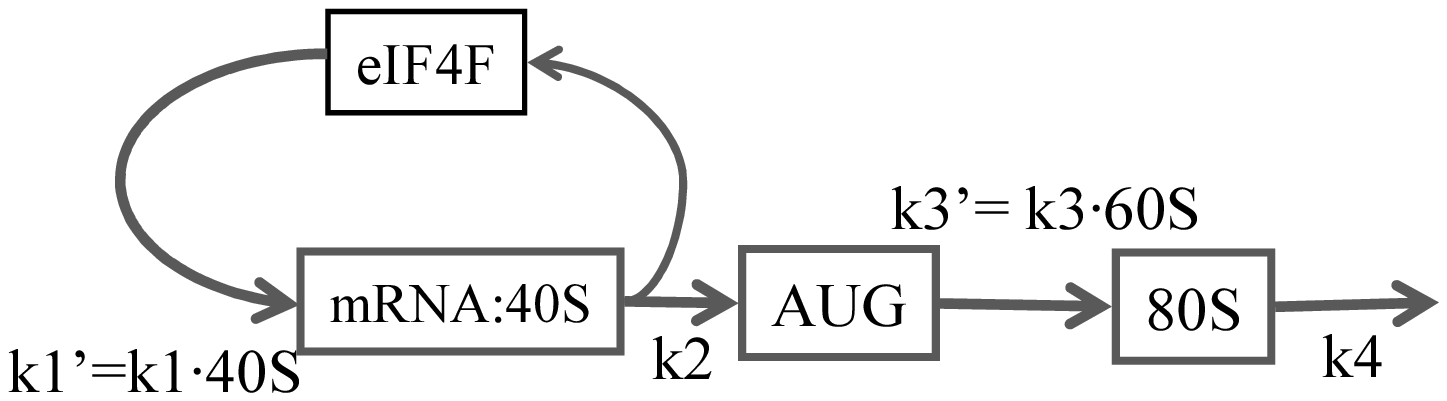}\hspace{0.7cm}\includegraphics[width=5cm,height=3cm]{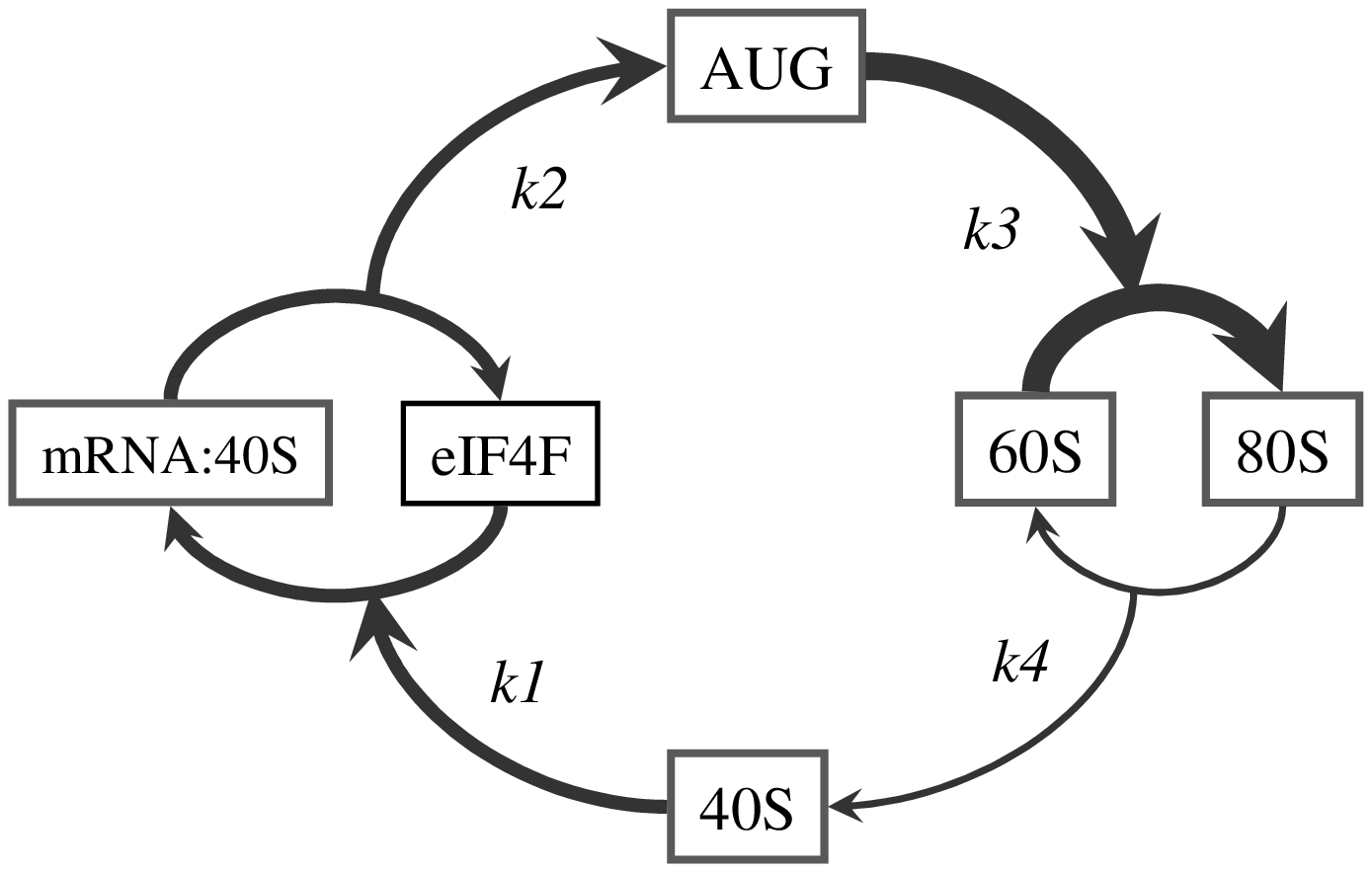}\hspace{0.7cm}
\includegraphics[width=5cm,height=3cm]{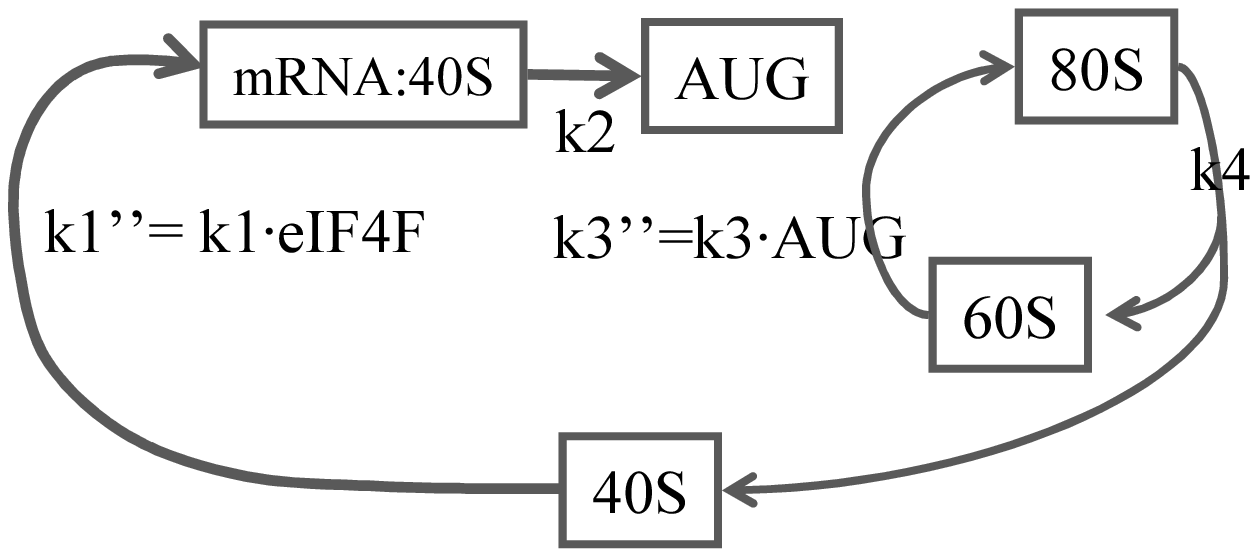}}
\centerline{Stage 1\hspace{5cm}Stage 2\hspace{5cm}Stage 3}
\caption{\label{Parker2DominantSubsystems} Dominant systems for
three stages of relaxation of the model
(\ref{ParkerNonlinearEquations}). Stage 1) The dominant system is
pseudo-linear network of reactions. Stage 2) The dominant system
is a quasi-steady state approximation, where one supposes that the
fluxes in two network cycles are balanced. Stage 3) The dominant
system is pseudo-linear network of reactions. }
\end{figure}


\begin{equation}
\left\{ \begin{split}
   &\frac{d\thinspace [eIF4F]}{dt} = -k_1'[eIF4F]+k_2[mRNA:40S] \\
   &\frac{d\thinspace [mRNA:40S]}{dt} = k_1'[eIF4F]-k_2[mRNA:40S]    \\
   &\frac{d\thinspace [AUG]}{dt} = k_2[mRNA:40S]-k_3'[AUG] \\
   &\frac{d\thinspace [80S]}{dt} = k_3[AUG]~[60S]-k_4[80S]\\
   \end{split}
\right. \label{ParkerNonlinearStage1}
\end{equation}

\noindent where $k_1'=k_1\cdot [40S]$, $k_3'=k_3\cdot [60S]$ and
we consider that at this stage the changes of 40S and 60S are
relatively slow. This system has simple approximate solution,
taking into account constraints on the parameters $k_2\ll
k_1',k_3'$; $k_4\ll k_1',k_3',k_2$, also assuming $k_2\ll
|k_1'-k_3'|$, and for the initial condition

\begin{equation}
\left[
\begin{split}
&[eIF4F] \\ &[mRNA:40S] \\ &[AUG]\\ &[80S]\\
\end{split}
\right]_0 = \left[
\begin{split}
[eIF4F]_0 \\ 0\\ 0\\ 0 \\
\end{split}
\right] \label{ParkerNonLinearInitialConditionStage1} \end{equation}

\noindent we have

\begin{equation}
\begin{split}&\left[
\begin{split}
&[eIF4F](t) \\ &[mRNA:40S](t)\\ &[AUG](t) \\ &[80S] \\
\end{split}
\right] \\  &= k_2\cdot [eIF4F]_0 \left( \left[
\begin{split}
&1/k_1' \\ &1/k_2 \\ &1/k_3' \\ &1/k_4
\end{split}
\right]-\frac{1}{k_3'} \left[
\begin{split}
&0\\ &0 \\ &1 \\ &-1
\end{split}
\right]e^{-k_3't}+\frac{1}{k_2}\left[
\begin{split}
&1\\&-1 \\ &0 \\ &0
\end{split}
\right]e^{-k_1't} - \frac{1}{k_4}\left[
\begin{split}
&0\\&0 \\ &0 \\ &1
\end{split}
\right]e^{-k_4t} \right  )
\label{ParkerLinearSolutionStage1}\end{split}
\end{equation}

From this solution, one can conclude that the relaxation of this model goes at several
 time scales (very rapid $\sim min(1/k_3',1/k_1')$ and slow $\sim 1/k_4$) and that
when eIF4F, mRNA:40S and AUG already reached their
quasiequilibrium values, [80S] continues to grow. This corresponds
to the quasiequilibrium approximation asymptotic (see section
\ref{asymptoticApprox}). At some point 80S will reach such a value
that it would be not possible to consider 60S constant: otherwise
the conservation law (\ref{ParkerNonlinearConservationLaw60S})
will be violated. This will happen when $[80S]<<[60S]$ condition
is not satisfied anymore, i.e., following our convention to
consider ''much smaller`` as difference in one order of magnitude,
at $t'\sim \frac{[60S]_0}{10\cdot [eIF4F]_0\cdot k_2}$. The same
consideration is applicable for another conservation law
(\ref{ParkerNonlinearConservationLaw40S}) in  which $[80S]$ is
included, but from the time point $t''\sim
\frac{\frac{[40S]_0}{10\cdot [eIF4F]_0}-1}{k_2}$. From $[40S]_0\gg
[eIF4F]_0$ and $[40S]_0>[60S]_0$ we have $min(t',t'')=t'$. This
means that the parameters $[40S]$, $[60S]$ of the (local) steady
states for $[eIF4F[$ and $[AUG]$ should slowly (at the same rate
as $[80S]$) change from the time point $t'$ (variable $[mRNA:40S]$
does not change because its local steady state does not depend on
$[40S]$, $[60S]$). In other words, after $t=t'$ the Stage 1
solution (\ref{ParkerLinearSolutionStage1}) should be prolonged as

\begin{equation}
\begin{split}
&[80S](t)=\frac{k_2\cdot [eIF4F]_0}{k_4}\left(1-e^{-k_4t}\right),
[40S](t)=[40S]_0-[80S](t), [60S](t)=[60S]_0-[80S](t), \\
&[eIF4F](t)=\frac{k_2\cdot [eIF4F]_0}{k_1([40S]_0-[80S](t))},
[mRNA:40S](t)=[eIF4F]_0, [AUG](t)=\frac{k_2\cdot
[eIF4F]_0}{k_3([60S]_0-[80S](t))}
\end{split}
  \label{ParkerStage1SSCorrection}.
\end{equation}

From these equations, one can determine the effective duration of
the Stage 1: by definition, it will be finished when one of the
two conditions ($[40S]\gg [eIF4F]$, $[60S]\gg [AUG]$) will be
violated, which happens at times $\sim\frac{[40S]_0}{k_2\cdot
[eIF4F]_0}$ and $\sim\frac{[60S]_0}{k_2\cdot [eIF4F]_0}$
correspondingly, hence, the second condition will be violated
first (from $[60S]_0<[40S]_0$).

\subsubsection{Stage 2: first stage of protein elongation,
initiation is still rapid}

The Stage 2 is characterized by conditions $[eIF4F] \ll [40S]$,
$[60S] \ll [AUG]$. This fact can be used for deriving the
quasi-steady state approximation: we assume that the reaction
fluxes in two network cycles (R1-R2 and R3-R4) are independently
balanced:

\begin{equation}
\begin{array}{lllll}
k_1[40S]\cdot [eIF4F] = k_2[mRNA:40S], k_3[AUG]\cdot [60S] =
k_4[80S].
\end{array}
  \label{Parker2BalanceEquations}
\end{equation}

Then (\ref{ParkerNonlinearEquations}) is simplified and, using the
conservation laws, we have a single equation on $[40S]$:

\begin{equation}
\begin{array}{lllll}
\frac{d\thinspace [40S](t)}{dt} = \frac{-k_2[eIF4F]_0\cdot
[40S](t)}{\frac{k_2}{k_1}+[40S](t)}+\frac{k_4[60S]_0(A-[40S](t))}{\frac{k_4}{k_3}+A-[40S](t)},
\end{array}
  \label{Parker240S}
\end{equation}

\noindent where $A = [40S](t)+[AUG](t)$ is a constant quantity
conserved accordingly to the quasi-steady state approximation.
Equation (\ref{Parker240S}) can be already integrated but let us
further simplify it for our analysis. Having in mind $k_4\ll k_3$
and assuming that at the beginning of the Stage 2 $[AUG]\gg
\frac{k4}{k3}$, we can simplify (\ref{Parker240S}) to

\begin{equation}
\begin{array}{lllll}
\frac{d\thinspace [40S](t)}{dt} = \frac{-k_2[eIF4F]_0\cdot
[40S](t)}{\frac{k_2}{k_1}+[40S](t)}+k_4[60S]_0,
\end{array}
  \label{Parker240S1}
\end{equation}

\noindent and, further, assuming that at the beginning of the
Stage 2 we have $[40S]\gg [AUG]$ let us approximate the right-hand
side of the equation by a piecewise-linear function

\begin{equation}
\frac{d\thinspace [40S](t)}{dt} \approx
\left\{\begin{array}{lllll}
-k_2[eIF4F]_0\frac{\frac{k_2}{k_1}+[40S]|_{t=t''}}{3\frac{k_2}{k_1}+[40S]|_{t=t''}}
+k_4[60S]_0,\ if\ [40S](t)>\frac{k_2}{k_1},
\\
-\frac{k_1[eIF4F]_0}{2}\cdot [40S](t) +k_4[60S]_0,\ if\
[40S](t)<\frac{k_2}{k_1}  ,
\end{array} \right.
  \label{Parker240SRHS}
\end{equation}

\noindent where $[40S]|_{t=t''}$ is the amount of $40S$ at the
beginning of the Stage 2. Then the descent of $[40S](t)$ can be
separated into linear and exponential phases:

\begin{equation}
[40S](t) \approx \left\{
\begin{split}
&[40S]|_{t=t''}-K_1\cdot (t-t''),\ if\ t<t''+\tau_{2} \\
&[40S]_{s2}-([40S]_{s2}-[40S]|_{t=t''})\exp\left(-K_2(t-t''-\tau_{2})\right),
\ if \ t>t''+\tau_{2}
\end{split} \right.
  \label{Parker240SSolution} ,
\end{equation}

\noindent where $K_1, K_2$ are linear and exponential slopes and
$[40S]_{s2}$ is the quasi-steady state value of $[40S]$ at the end
of the Stage 2:

\begin{equation}
\begin{split}
&K_1 =
k_2eIF4F_0\frac{\frac{k_2}{k_1}+[40S]_{t=t''}}{3\frac{k_2}{k_1}+[40S]|_{t=t''}}
-k_4[60S]_0 \\ &K_2 = \frac{k_1[eIF4F]_0}{2}
\\ &[40S]_{s2} =
\frac{k_2}{k_1}\frac{1}{\frac{k_2[eIF4F]_0}{k_4[60S]_0}-1}\\
&\tau_{2} = \frac{[40S]|_{t=t_2}-\frac{k_2}{k_1}}{K_1}
\end{split}
\end{equation}

Other dynamic variables are expressed through $[40S](t)$ as

\begin{equation}
\begin{split}
&[eIF4F](t) =
[eIF4F]_0\frac{\frac{k_2}{k_1}}{\frac{k_2}{k_1}+[40S](t)}
\\
&[mRNA:40S](t) = [eIF4F]_0\cdot
\frac{[40S](t)}{\frac{k_2}{k_1}+[40S](t)}
\\
&[AUG](t) =
[40S]|_{t=t''}+\frac{k_2[eIF4F]_0}{k_3([60S]_0-[40S]_0+[40S]|_{t=t''})}-[40S](t)\\
&[60S](t) =
[60S]_0\frac{\frac{k_4}{k_3}}{\frac{k_4}{k_3}+[AUG](t)}\\
&[80S]_0(t) = [60S]_0\frac{[AUG](t)}{\frac{k_4}{k_3}+[AUG](t)}
\end{split}
  \label{Parker240SSolution1} .
\end{equation}

At some point, the amount of free small ribosomal subunit 40S,
which is abundant at the beginning of the Stage 2, will not be
sufficient to support rapid translation initiation. Then the
initiation factor $eIF4F$ will not be the limiting factor in the
initiation and the condition $[40S]\gg [eIF4F]$ will be violated.
We can estimate this time as $t'''=\frac{[40S]_0}{k_2\cdot
[eIF4F]_0}$.

\subsubsection{Stage 3: steady protein elongation, speed of initiation equals to speed of elongation}

During the Stage 3 all fluxes in the network become balanced and
the translation arrives at the steady state. From
Fig.~\ref{Parker2DominantSubsystems} it is clear that the
relaxation goes independently in the cycle $R3-R4$, where the
relaxation equations are simply

\begin{equation}
[60S](t) =
[60S]_s+([60S]|{t=t'''}-[60S]_s)e^{-(k_4+k_3[AUG]|{t=t'''})(t-t''')},
[80S](t) = [60S]_0-[60S](t).
\end{equation}

\noindent where $t'''$ is the time when the Stage 3 of the
relaxation starts. This relaxation goes relatively fast, since
$k_3[AUG]|_{t=t'''}$ is relatively big. So, during the Stage 3,
one can consider the cycle $R1-R4$ equilibrated, with
$[80S]=[80]_s$, $[60S]=[60]_s$ values.

Hence, the relaxation during the Stage 3 consists in
redistributing concentrations of 40S and mRNA:40S to their steady
states in a linear chain of reactions $R1-R2$ (the value of
$[AUG]$ is relatively big and can be adjusted from the
conservation law (\ref{ParkerNonlinearConservationLaw40S})). Using
the pseudo-linear approximation of this stage (see
Fig.\ref{Parker2DominantSubsystems}), we can easily write down the
corresponding approximate relaxation equations:

\begin{equation}
\begin{split}
&[mRNA:40S](t)= \\
&[mRNA:40]_s(1-e^{-k_2t})+B(e^{-k_1[eIF4F]|_{t=t'''}(t-t''')}-e^{-k_2(t-t''')})+[mRNA:40S]|_{t=t'''}e^{-k_2(t-t''')},
\\&[40S](t)=[40S]_s+Be^{-k_2(t-t''')}
\end{split}
\end{equation}

\noindent where
$B=(1-\frac{k_2}{k_1[eIF4F]'''})([40]_s-[40S]|_{t=t'''})$.
$[40]_s$ and $[mRNA:40S]_s$ are the steady-state values of the
corresponding variables, see (\ref{ParkerNonlinearSteadyState}).
The values $[60S]|_{t=t'''}$, $[eIF4F]|_{t=t'''}$,
$[AUG]|_{t=t'''}$ and $[mRNA:40S]|_{t=t'''}$ can be estimated from
(\ref{Parker240SSolution1}), using the $[40S]|_{t=t'''}$ value.
The relaxation time at this stage equals $$\tau_3 =
\max(\frac{1}{k_1[eIF4F]|_{t=t'''}},\frac{1}{k_2},\frac{1}{k_4+k_3[AUG]|_{t=t'''}})$$

The solution for the Stage 3 can be further simplified if $k_2\ll
k_1[eIF4F]|_{t=t'''}$ or $k_2\gg k_1[eIF4F]|_{t=t'''}$.

\subsubsection{Transitions between stages}

Along the trajectory of the dynamical system
(\ref{ParkerNonlinearEquations}) there are three dominant system
each one transforming into an other. At the transition between
stages, two neighbor dominant systems are united and then split.
Theoretically, there might be situations when the system can stay
in these transition zones for long periods of time, even
infinitely. However, in the model (\ref{ParkerNonlinearEquations})
this is not the case: the trajectory rapidly passes through the
transition stages and jumps into the next dominant system
approximation.

Three dominant approximations can be glued, using the
concentration values at the times of the switching of dominant
approximation as initial values for the next stage. Note that the
Stage 2 has essentially one degree of freedom since it can be
approximated by a single equation (\ref{Parker240S}). Hence, one
should only know one initial value $[40S]|_{t=t''}$ to glue the
Stages 1 and 2. The same is applied to the gluing of Stages 2 and
3, since in the end of Stage 2 all variable values are determined
by the value of $[40S]|_{t=t'''}$.

\subsubsection{Case of always limiting initiation}

As it follows from our analysis, the most critical parameter of
the non-linear protein translation model is the ratio $\beta =
\frac{k_2[eIF4F]_0}{k_4[60S]_0}$. Above we have considered the
case $\beta>1$ which is characterized by a switch of the limiting
factor in the initiation (from $eIF4F$ at the Stages 1 and 2 to
$40S$ at the Stage 3).

In the case $\beta<1$ the dynamics becomes simpler and consists of
one single stage: relaxation accordingly to
(\ref{ParkerLinearSolutionStage1}) and further with correction
(\ref{ParkerStage1SSCorrection}) with the relaxation time $\sim
\frac{1}{k_4}$ (the quasiequilibrium approximation corresponding
to the Stage 1 works well for the whole translation process). The
reason for this is that if the initiation is not efficient then
the system is never in the situation of the Stage 2 conditions
when the cycle R1-R2 is balanced with much bigger flux than the
cycle R3-R4. This approximation is the more exact the smaller
$\beta$ value, however, the value of $\beta$ should not be
necessary very small. For example, for the default parameter
values of the model $\beta=0.48$, and it well reproduces the
dynamics (see Fig.~\ref{Parker2SolutionApp}c-d). From numerical
experiments one can see that even for $\beta=0.95$ the dynamics is
qualitatively well reproduced.

To model the A-cap structure effect with very weak capacity for
initiation (assembly of the initiation factors and 40S subunit),
we should also consider the case

\begin{equation}
k_1 \ll k_4 \ll k_2 \ll k_3,
\end{equation}

\noindent for which the solution derived above is not directly
applicable. However, the analytical calculations in this case can
be performed in the same fashion as above. The detailed derivation
of the solution is given in the Supplementary materials. The
effect of putting $k_1$ very small on the steady state protein
synthesis and the relaxation time is shown on
Fig.~\ref{Parker2SolutionApp}.

In a similar way all possible solutions of the equations
(\ref{ParkerNonlinearEquations}) with very strong inhibitory
effect of microRNA on a particular translation step can be
derived. These solutions will describe the situation when the
effect of microRNA is so strong that it changes the dominant
system (limiting place of the network) by violating the initial
constraints (\ref{ParkerNonlinearParameterOrders}) on the
parameters (for example, by making $k_3$ smaller than other
$k_i$s). Such possibility exists, however, it can require too
strong (non-physiological) effect of microRNA-dependent
translation inhibition.

\subsubsection{Effect of microRNA on the translation dynamics}


\begin{figure}
\centerline{a)\includegraphics[width=9cm,
height=7cm]{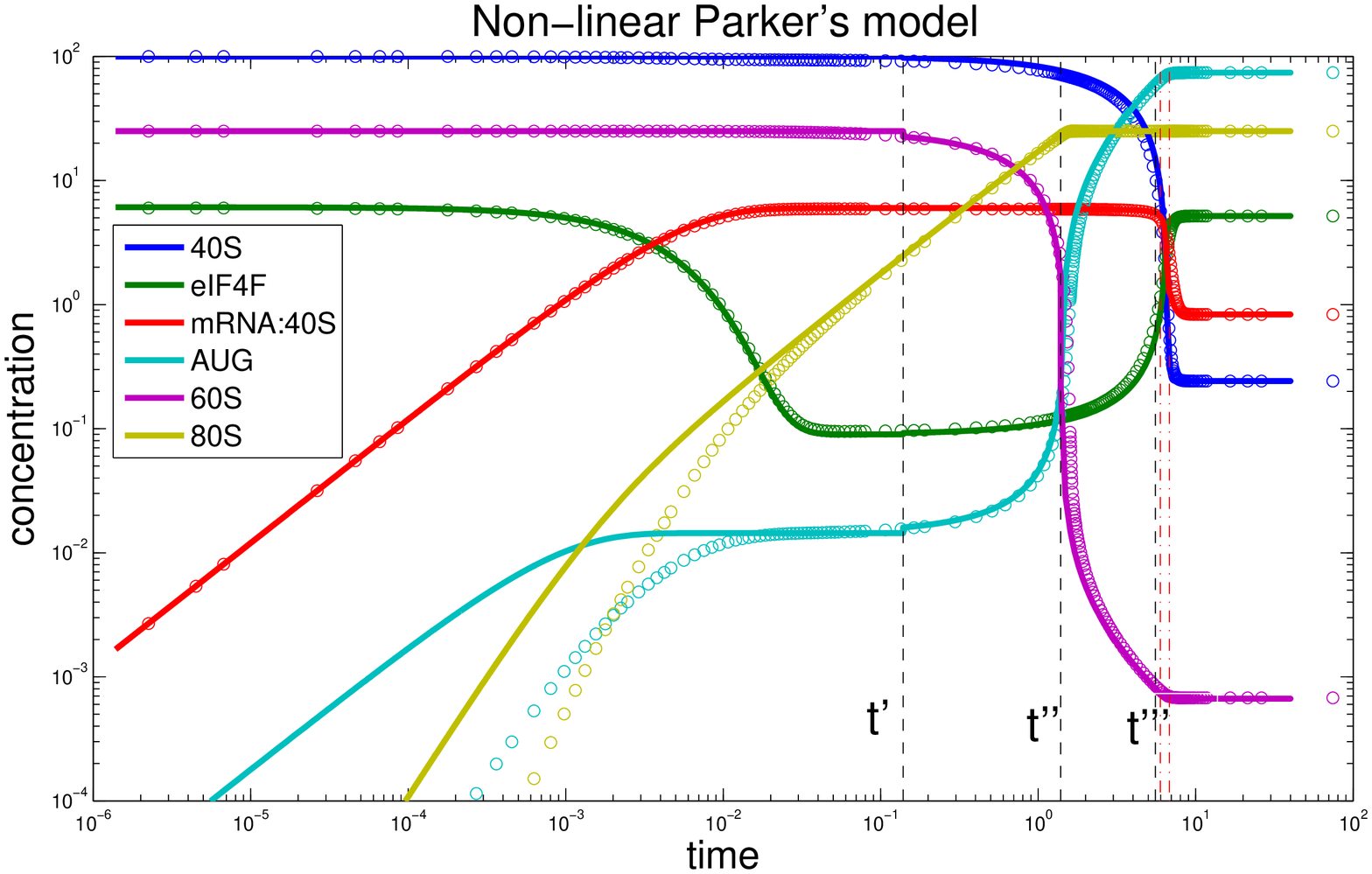}b)\includegraphics[width=9cm,
height=7cm]{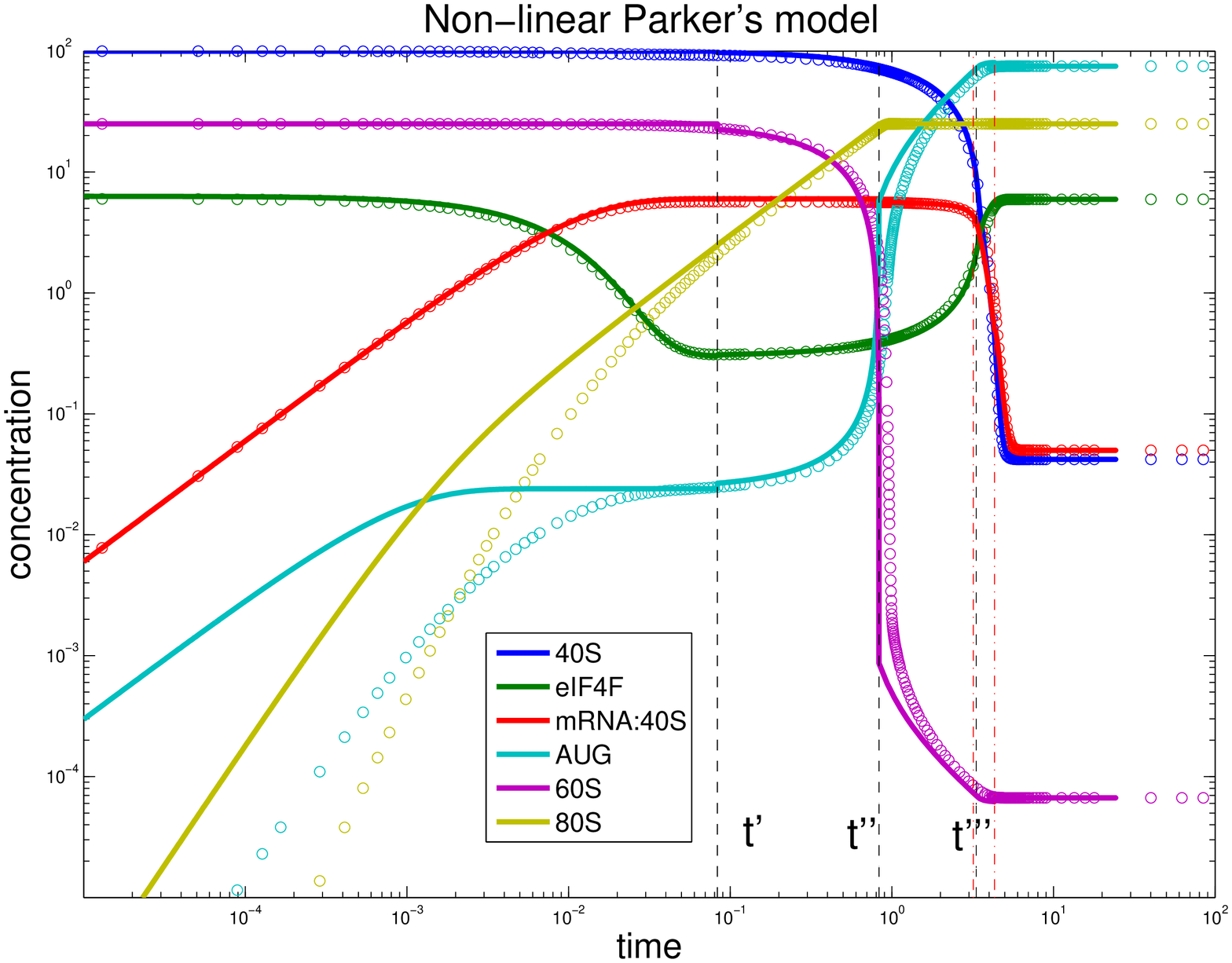}} \centerline{
c)\includegraphics[width=9cm,
height=7cm]{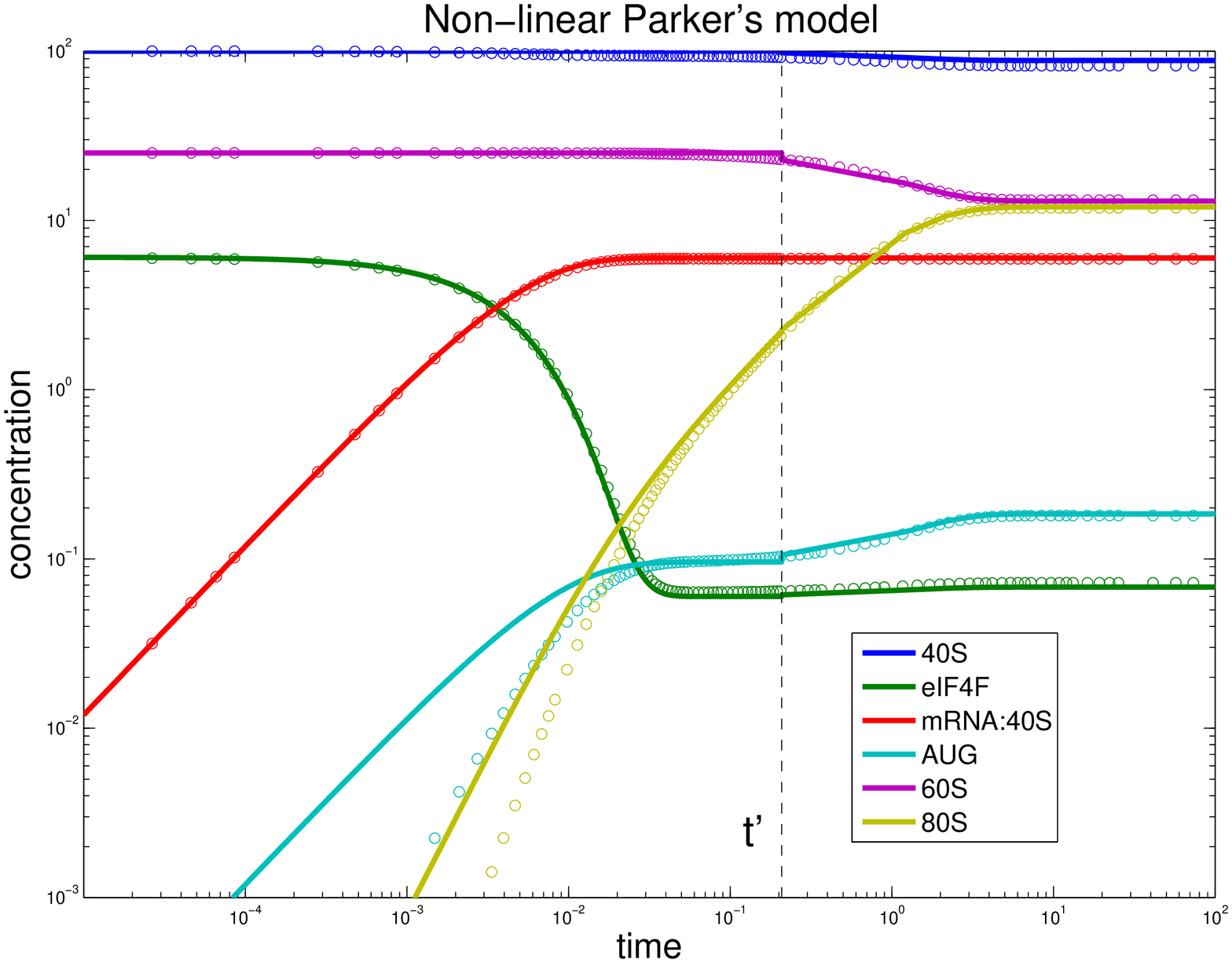}d)\includegraphics[width=9cm,
height=7cm]{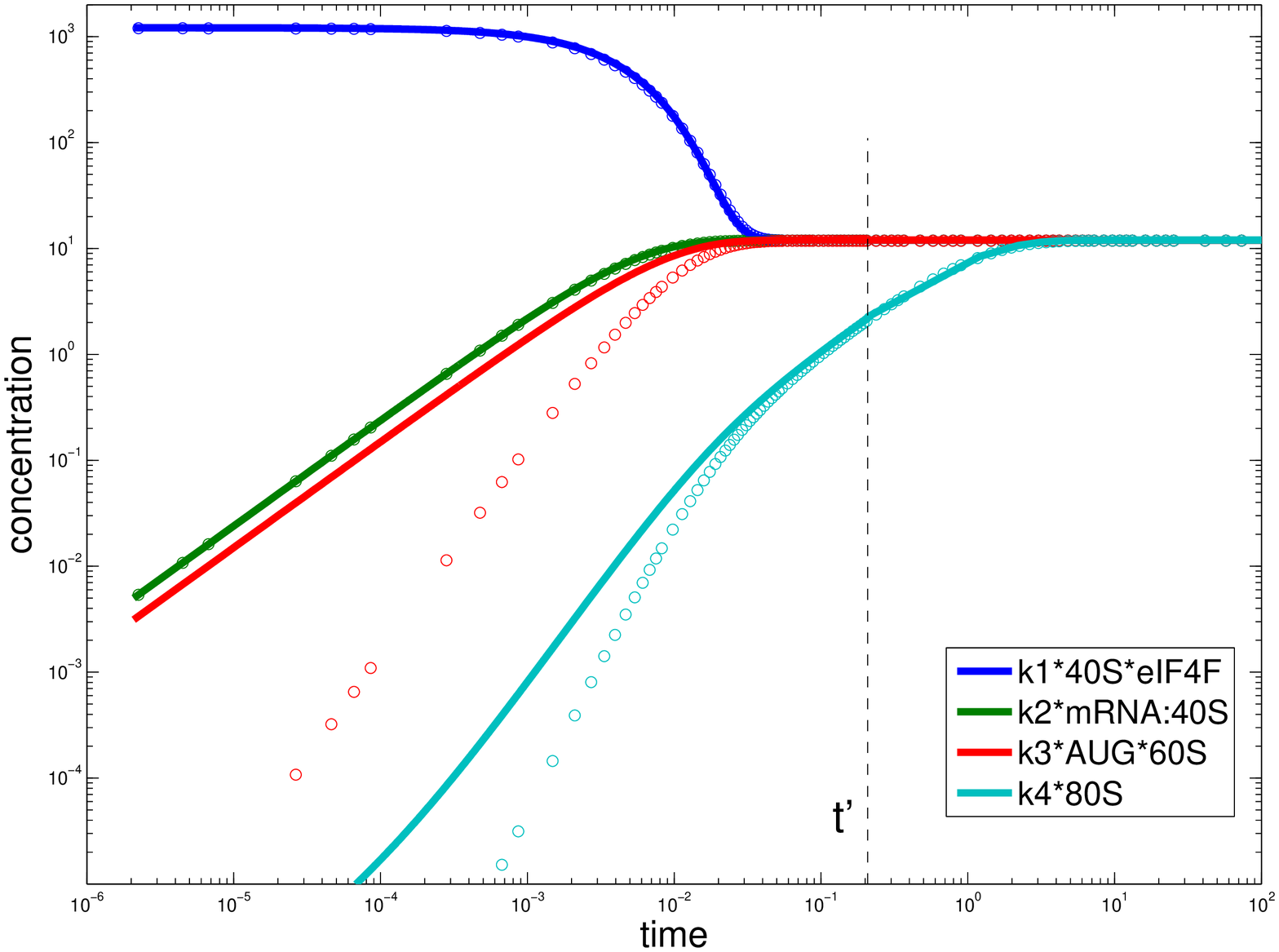}}
\caption{\label{Parker2SolutionApp} Examples of the exact
(circles) and approximate (solid lines) solutions of the
non-linear protein translation model. a) For our set of parameters
$k_1=2$, $k_2=3$, $k_3=50$, $k_4=0.1$; b) For parameters $k_1=1$,
$k_2=5$, $k_3=50$, $k_4=0.01$; c) For the set of parameters from
\cite{Nissan2008}; d) Reaction fluxes for the set of parameters
c). Dashed black vertical lines denote evaluated transition points
between the dynamics stages. Dashed red vertical points denote the
time points where $[40S](t)=10\cdot [eIF4F](t)$ and
$[40S](t)=[eIF4F](t)/10$ respectively. }
\end{figure}

As a result of the above analysis, we can assemble an approximate
solution of the non-linear system under assumptions
(\ref{ParkerNonlinearParameterOrders}) about the parameters. An
example of the approximate solution is given on
Fig.~\ref{Parker2SolutionApp}. The advantage of such a
semi-analytical solution is that one can predict the effect of
changing the system parameters. For example, on
Fig.~\ref{Parker2SolutionApp}b the solution is compared to an
exact numerical one, where the parameters have been changed but
still obey the initial constraints
(\ref{ParkerNonlinearParameterOrders}).

One of the obvious predictions is that the dynamics of the system
is not sensitive to variations of $k_3$, so if microRNA acts on
the translation stage controlled by $k_3$ then no microRNA effect
could be observed looking at the system dynamics (being the
fastest one, $k_3$ is not a critical parameter in any scenario).

\begin{figure}
.\hspace{1cm}{\bf Wild-type cap, inefficient
initiation\hspace{2.5cm}Wild-type cap, efficient initiation}\\
\includegraphics[width=4.5cm,height=4cm]{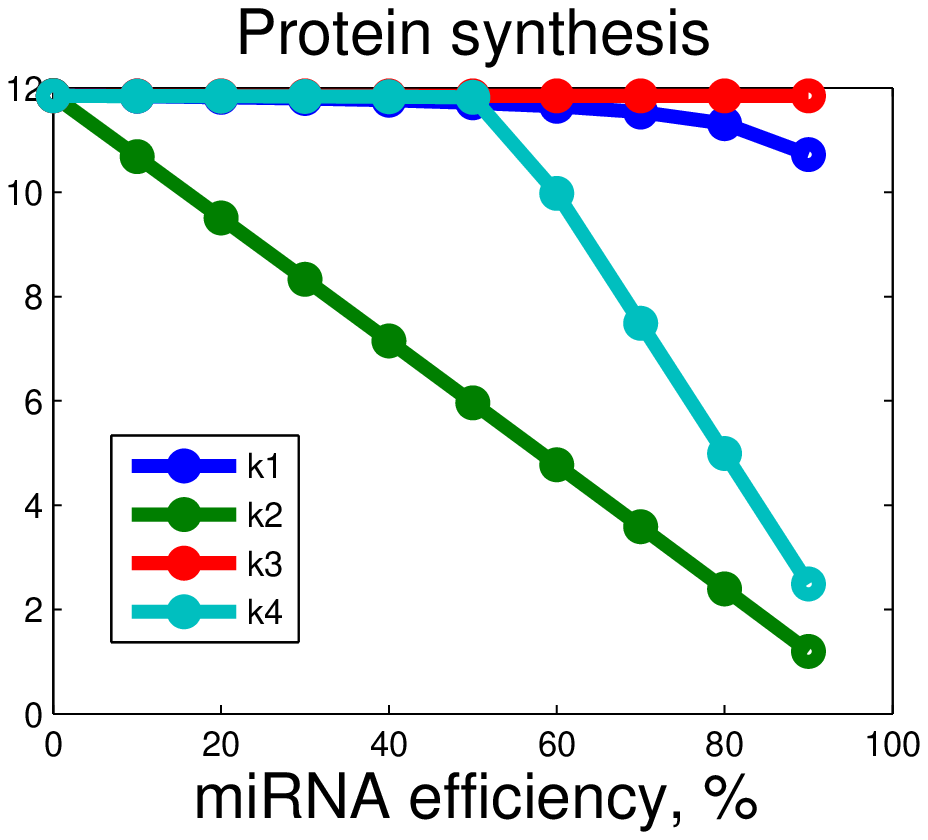}\includegraphics[width=4.5cm,height=4cm]{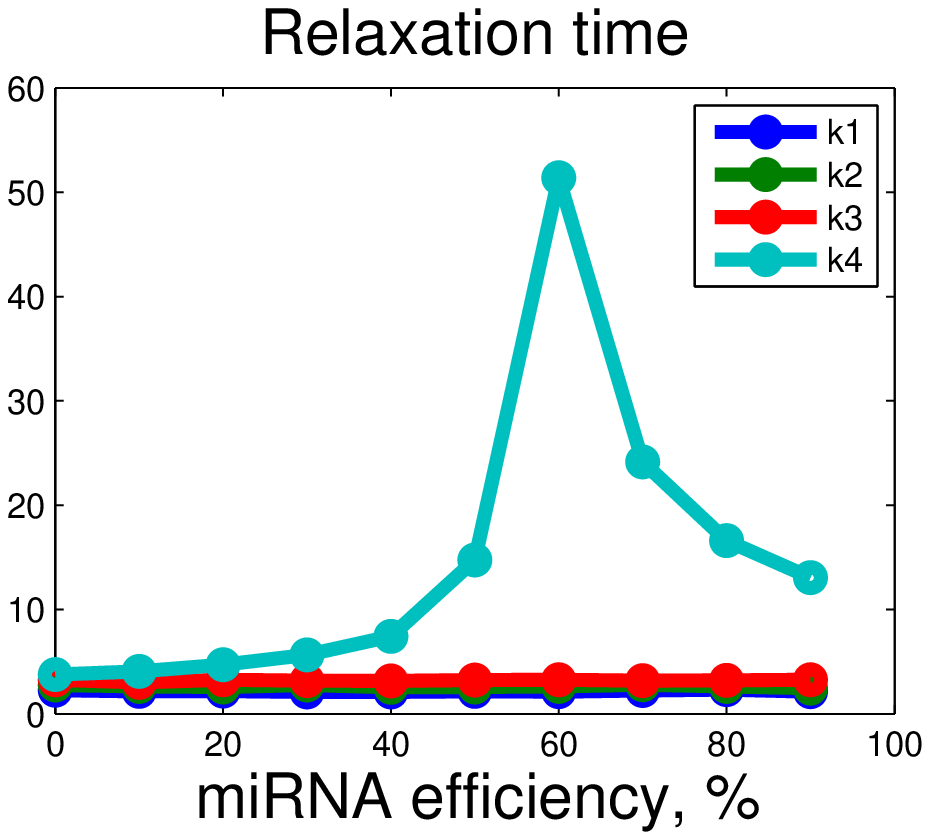}
\includegraphics[width=4.5cm,height=4cm]{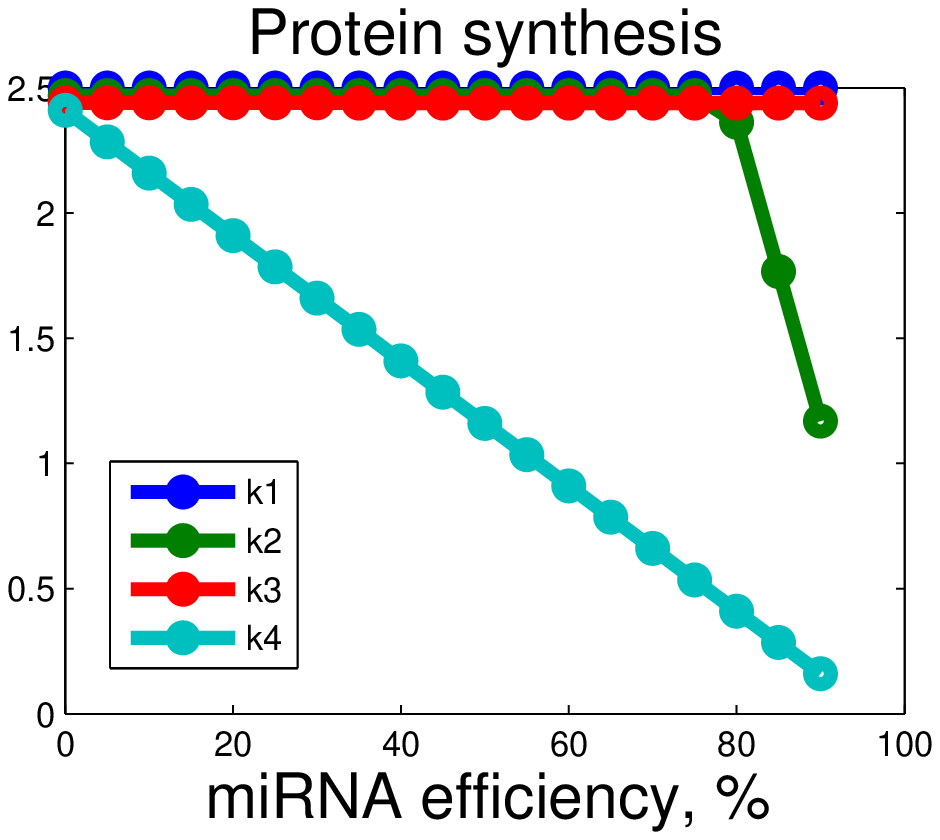}\includegraphics[width=4.5cm,height=4cm]{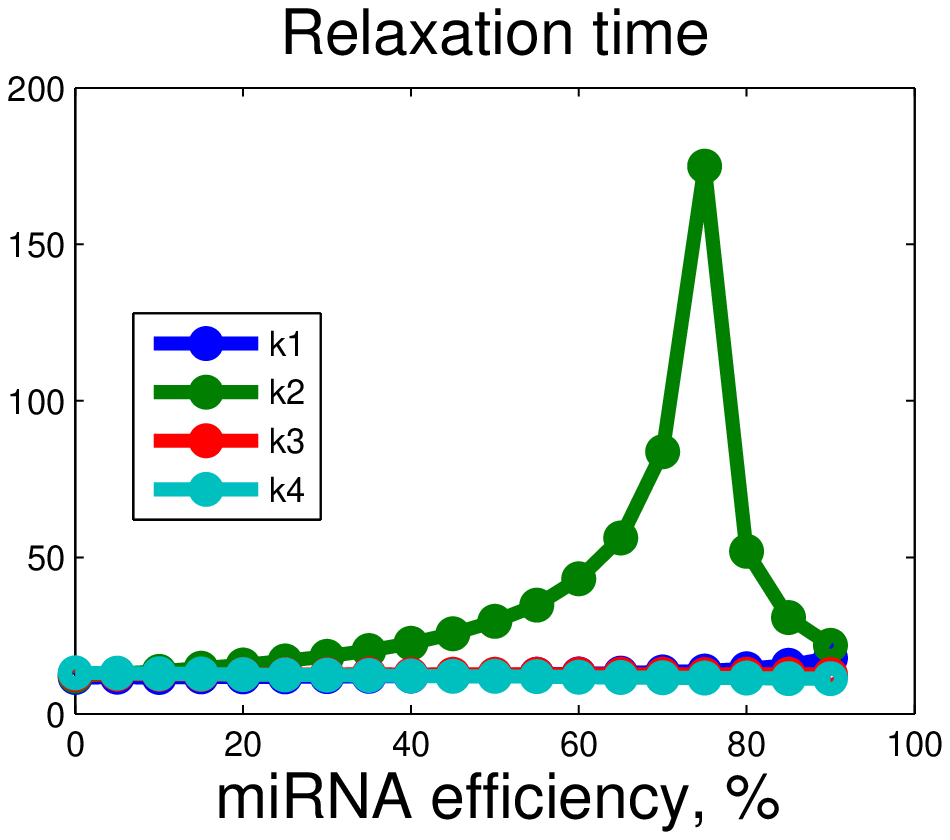}\\
.\hspace{2cm}A)\hspace{4.5cm}B)\hspace{4.5cm}C)\hspace{4.5cm}D)\\
.\hspace{2cm}{\bf A-cap, inefficient
initiation\hspace{3.5cm}A-cap, efficient initiation}\\
\includegraphics[width=4.5cm,height=4cm]{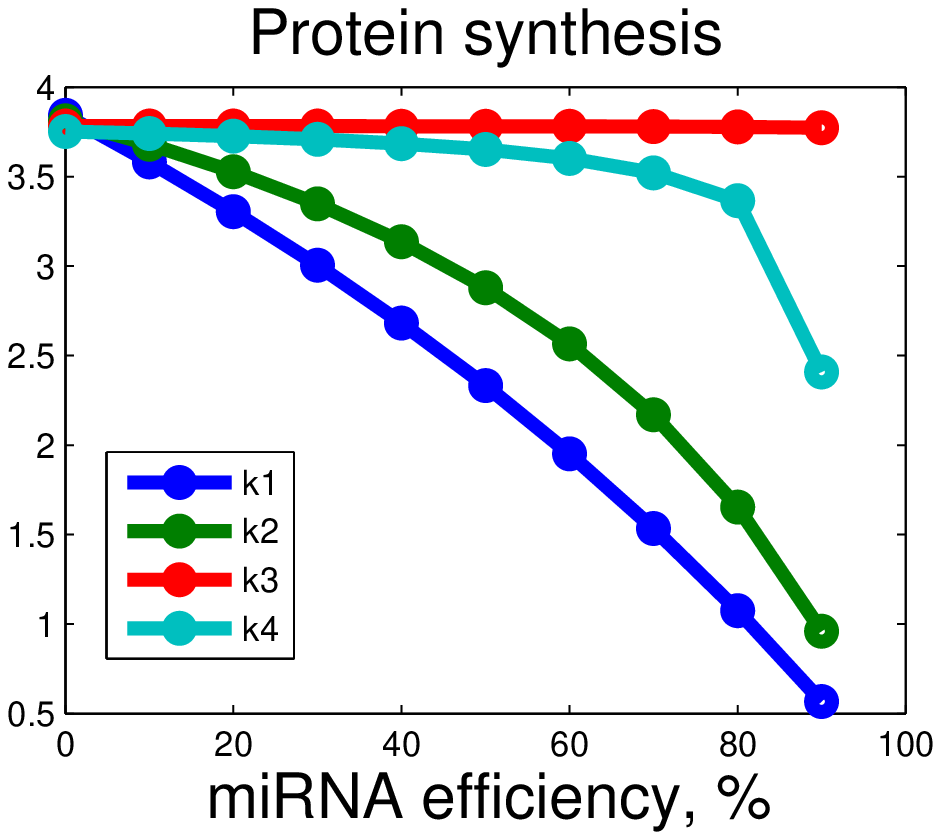}\includegraphics[width=4.5cm,height=4cm]{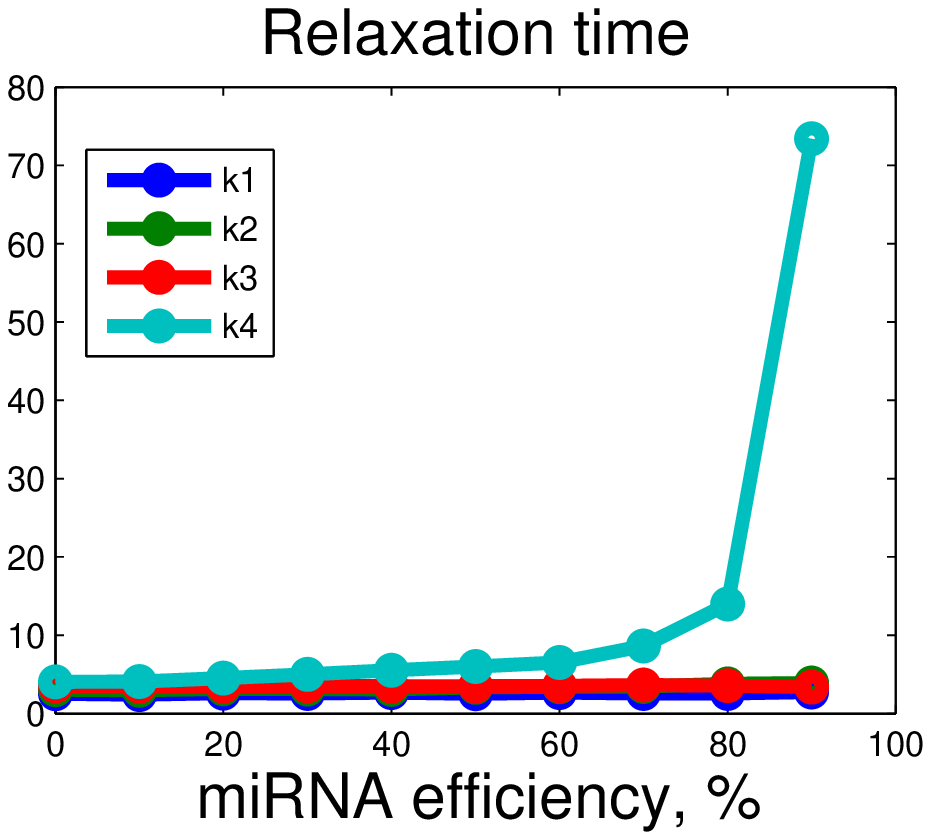}
\includegraphics[width=4.5cm,height=4cm]{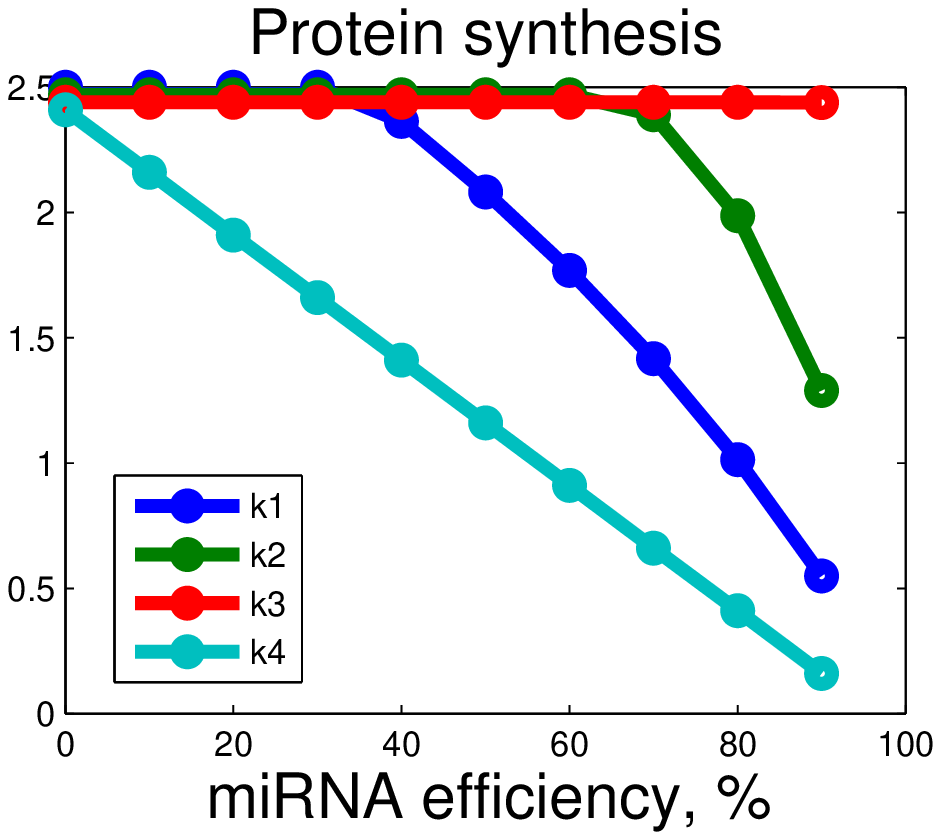}\includegraphics[width=4.5cm,height=4cm]{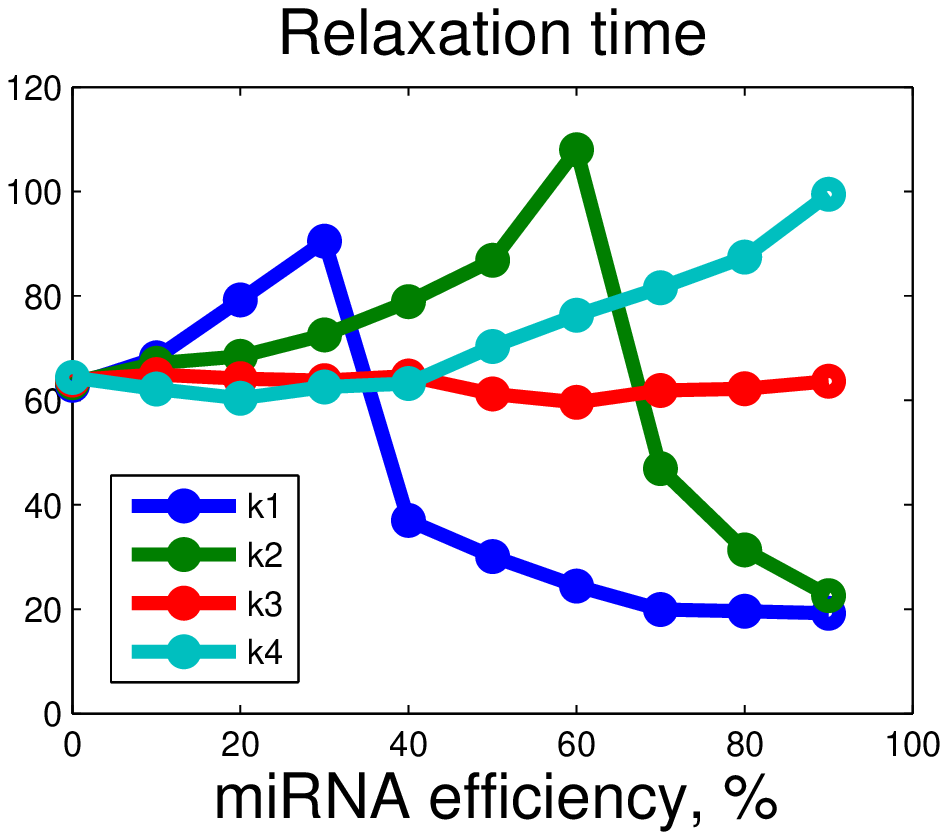}\\
.\hspace{2cm}E)\hspace{4.5cm}F)\hspace{4.5cm}G)\hspace{4.5cm}H)
\caption{\label{Parker2microRNAeffect} Effect of decreasing some
model parameters mimicking different mechanisms of miRNA action on
translation. Relaxation time here is defined as the latest time at
which any chemical species in the model differs from its final
steady state by 10\%. A) and B) correspond to the scenario with
''inefficient`` initiation, with use of the model parameters
proposed in \cite{Nissan2008} ($k_1=k_2=2$, $k_3=5$, $k_4=1$,
$[eIF4F]_0=6$, $[60S]_0=25$, $[40S]_0=100$), which gives
$\beta=0.48<1$. C) and D) correpond to the scenario with
''efficient`` initiation, with our choice of parameters ($k_1=2$,
$k_2=3$, $k_3=50$, $k_4=0.1$, $[eIF4F]_0=6$, $[60S]_0=25$,
$[40S]_0=100$), which gives $\beta=7.2>1$. The absciss value
indicates the degree of inhibition (decreasing) of a parameter.
E-H) same as A-D) but for a modified cap structure, modeled by
reduced $k_1$ parameter: $k_1=0.01$ for these curves, the other
parameters are the same as on A-D) correspondingly. }
\end{figure}

If microRNA acts on the translation stage controlled by $k_4$ (for
example, by ribosome stalling mechanism) then we should consider
two cases of efficient ($\beta>1$) and inefficient ($\beta<1$)
initiation. In the first case the steady state protein synthesis
rate is controlled by $k_4$ (as the slowest, limiting step) and
any effect on $k_4$ would lead to the proportional change in the
steady state of protein production. By contrast, in the case of
inefficient initiation, the steady state protein synthesis is not
affected by $k_4$. Instead, the relaxation time is affected, being
$\sim \frac{1}{k_4}$. However, diminishing $k_4$ increases the
$\beta$ parameter, hence, this changes ``inefficient initiation''
scenario for the opposite, hence, making $k_4$ critical for the
steady state protein synthesis anyway when $k_4$ becomes smaller
than $\frac{k_2[eIF4F]_0}{[60S]_0}$. For example, for the default
parameters of the model, decreasing $k_4$ value first leads to no
change in the steady state rate of protein synthesis but increases
the relaxation time and, second, after the threshold value
$\frac{k_2[eIF4F]_0}{[60S]_0}$ starts to affect the steady state
protein synthesis rate directly (see
Fig.~\ref{Parker2microRNAeffect}A). This is in contradiction to
the message from \cite{Nissan2008} that the change in $k_4$ by
several orders of magnitude does not change the steady state rate
of protein synthesis.

Analogously, decreasing the value of $k_2$ can convert the
``efficient'' initiation scenario into the opposite after the
threshold value $\frac{k_4[60S]_0}{[eIF4F]_0}$. We can
recapitulate the effect of decreasing $k_2$ in the following way.
1) in the case of efficient initiation $k_2$ does not affect the
steady state protein synthesis rate up to the threshold value
after which it affects it in a proportional way. The relaxation
time drastically increases, because decreasing $k_2$ leads to
elongation of all dynamical stages durations (for example, we have
estimated the time of the end of the dynamical Stage 2 as
$t'''=\frac{[40S]_0}{k_2\cdot [eIF4F]_0}$). However, after the
threshold value the relaxation time decreases together with $k_2$,
quickly dropping to its unperturbed value. 2) in the case case of
''inefficient`` initiation the steady state protein synthesis rate
depends proportionally on the value of $k_2$
(\ref{ParkerNonLinearProtSynthesis}), while the relaxation time is
not affected (see Fig.~\ref{Parker2microRNAeffect}).

MicroRNA action on $k_1$ directly does not produce any strong
effect neither on the relaxation time nor on the steady state
protein synthesis rate. This is why in the original work
\cite{Nissan2008} cap-dependent mechanism of microRNA action was
taken into account through effective change of the $[eIF4F]_0$
value (total concentration of the translation initiation factors),
which is a critical parameter of the model (see
\ref{ParkerNonLinearProtSynthesis}).

The effect of microRNA on various mechanism and in various
experimental settings (excess or deficit of eIF4F, normal cap or
A-cap) is recapitulated in Table~\ref{tableNonLinear}. The
conclusion that can be made from this table is that all four
mechanism shows clearly different pattern of behaviour in various
experimental settings. From the simulations one can make a
conclusion that it is still not possible to distinguish between
the situation when microRNA does not have any effect on protein
translation and the situation when it acts on the step which is
neither rate limiting nor 'second rate limiting' in any
experimental setting ($k_3$ in our case). Nevertheless, if any
change in the steady-state protein synthesis or the relaxation
time is observed, theoretically, it will be possible to specify
the mechanism responsible for it.

\begin{table}
\caption{Modeling of four mechanisms of microRNA action in the
non-linear protein translation model. MicroRNA action effect is
described for the protein synthesis steady rate and the relaxation
time (see also Fig.~\ref{Parker2microRNAeffect}). It is assumed
that the ribosome assembly step in protein translation, described
by the $k_3$ rate constant, is not rate limiting
\label{tableNonLinear}} \centering{ \small{
\begin{tabular}{|c|p{2.5cm}|p{2.5cm}|c|p{2.5cm}|c|}\hline
{Observable value} & {Initiation($k_1$)} & {Step after
initiation($k_2$)} & {Ribosome assembly ($k_3$)} & {Elongation
($k_4$)}
\\\hline \multicolumn{5}{c}{{\bf Wild-type cap, inefficient initiation}}
\\ {\it Steady-state rate} & slightly decreases & decreases
& no change & decreases after threshold \\ {\it Relaxation time} &
no change & no change & no change & goes up and down
\\
\multicolumn{5}{c}{{\bf Wild-type cap, efficient initiation}}
\\ {\it Steady-state rate} & no change & slightly decreases \linebreak after
strong inhibition & no change & decreases \\ {\it Relaxation time}
& no change & goes up and down & no change & no change
\\
\multicolumn{5}{c}{{\bf A-cap, inefficient initiation}} \\ {\it
Steady-state rate} & decreases & decreases & no change & slighly
decreases after strong inhibition
\\ {\it Relaxation time} & no change & no change & no change & goes up and down\\
\multicolumn{5}{c}{{\bf A-cap, efficient initiation}} \\ {\it
Steady-state rate} & decreases after threshold & slightly
decreases after strong inhibition
 & no change & decreases
\\{\it
Relaxation time} & goes up and down & goes up and down
 & no change & increases \\ \hline
\end{tabular}}}
\end{table}

\section{Discussion}

The role of microRNA in gene expression regulation is discovered
and confirmed since ten years, however, there is still a lot of
controversial results regarding the role of concrete mechanisms of
microRNA-mediated protein synthesis respression. Some authors
argue that it is possible that the different modes of microRNA
action reflect different interpretations and experimental
approaches, but the possibility that microRNAs do indeed silence
gene expression via multiple mechanisms also exists. Finally,
microRNAs might silence gene expression by a common and unique
mechanism; and the multiple modes of action represent secondary
effects of this primary event \cite{Chekulaeva2009, Eulalio2008,
Filipowicz2008}.

The main reason for accepting a possible experimental bias could
be the studies in vitro, where conditions are strongly different
from situation in vivo. Indeed, inside the cell, mRNAs (microRNA
targets) exist as ribonucleoprotein particles or mRNPs, and
second, all proteins normally associated with mRNAs transcribed in
vivo are absent or at least much  different from that bound to the
same mRNA in an in vitro system or following the microRNAs
transfection into cultured cells. The fact that RNA-binding
proteins strongly influence the final outcome of microRNA
regulation is proved now by several studies
\cite{Bhattacharyya2006, George2006, Jing2005}.

The mathematical results provided in this paper suggests a
complementary view on the co-existence of multiple
microRNA-mediated mechanisms of translation repression.
Mathematical modeling suggests to us to ask a question: {\it if
multiple mechanisms act at the same time, would all of them
equally contribute to the final observable repression of protein
synthesis or its dynamics?} The dynamical limitation theory gives
an answer: {\it the effect of microRNA action will be observable
and measurable in two cases: 1) if it affects the dominant system
of the protein translationary machinery, or 2) if the effect of
microRNA action is so strong that it changes the limiting place
(the dominant system). }

In a limited sense, this means, in particular, that the protein
synthesis steady rate is determined by the limiting step in the
translation process and any effect of microRNA will be measurable
only if it affects the limiting step in translation, as it was
demonstrated in \cite{Nissan2008}. Due to the variety of external
conditions, cellular contexts and experimental systems the
limiting step in principle can be any in the sequence of events in
protein translation, hence, this or that microRNA mechanism can
become dominant in a concrete environment. However, when put on
the language of equations, the previous statement already becomes
non-trivial in the case of non-linear dynamical models of
translation (and even linear reaction networks with non-trivial
network structure). Our analysis demonstrates that the limiting
step in translation can change with time, depends on the initial
conditions and is not represented by a single reaction rate
constant but rather by some combination of several model
parameters. Methodology of dynamical limitation theory that we had
developed \cite{GorbanRadul2008, Gorban2009}, allows to deal with
these situations on a solid theoretical ground.

Furthermore, in the dynamical limitation theory, we generalize the
notion of the limiting step to the notion of dominant system, and
this gives us a possibility to consider not only the steady state
rate but also some dynamical features of the system under study.
One of the simplest measurable dynamical feature is the {\it
protein synthesis relaxation time}, i.e. the time needed for
protein synthesis to achieve its steady state rate. The general
idea of ``relaxation spectrometry'' goes back to the works of
Manfred Eigen, a Nobel laureate \cite{Eigen72} and is still
underestimated in systems biology. Calculation of the relaxation
time (or times) requires careful analysis of time scales in the
dynamical system, which is greatly facilitated by the recipes
proposed in \cite{Gorban2009,RadGorZinLil2008}. As we have
demonstrated in our semi-analytical solutions, measuring the
steady state rate and relaxation time at the same time allows to
detect which step is possibly affected by the action of microRNA
(resulting in effective slowing down of this step). To our
knowledge, this idea was never considered before in the studies of
microRNA-dependent expression regulation. The table
\ref{tableNonLinear} recapitulates predictions allowing to
discriminate a particular mechanism of microRNA action.

Thus, analysis of the transient dynamics gives enough information
to verify or reject a hypothesis about a particular molecular
mechanism of microRNA action on protein translation. For
multiscale systems only that action of microRNA is distinguishable
which affects the parameters of dominant system (critical
parameters), or changes the dominant system itself. Dominant
systems generalize and further develop the old and very popular
idea of limiting step. Algorithms for identifying dominant systems
in multiscale kinetic models are straightforward but not trivial
and depend only on the ordering of the model parameters but not on
their concrete values. Asymptotic approach to kinetic models of
biological networks suggest new directions of thinking on a
biological problem, making the mathematical model a useful tool
accompanying biological reasoning and allowing to put in order
diverse experimental observations.

\section{Author contributions}

AZ and AG have written the main body of the manuscript. NM, NN and
AHB provided the critical review of miRNA mechanisms and
contributed to writing the manuscript. AG and AZ developed the
mathematical methodology for identifying dominant systems. AZ
performed the analytical computations and numerical simulations.
All authors have participated in discussing the results and model
predictions.

\section{Acknowledgements}

We acknowledge support from Agence Nationale de la Recherche (ANR
CALAMAR project) and from the Projet Incitatif Collaboratif
``Bioinformatics and Biostatistics of Cancer" at Institut Curie.
AZ and EB are members of the team ''Systems Biology of Cancer``
Equipe labellis\'ee par la Ligue Nationale Contre le Cancer. We
thank Vitaly Volpert and Laurence Calzone for inspiring and useful
discussions.

\end{document}